\documentclass[aps,pra,twocolumn,longbibliography,superscriptaddress,10pt]{revtex4-2}
\usepackage{amsmath,amssymb,,epstopdf,amsthm,amssymb,amsthm,physics,bm,bbm,color,mathtools,anyfontsize,appendix,latexsym}	
\usepackage{latexsym}
\usepackage{amssymb}
\usepackage[colorlinks=true,linkcolor=blue,citecolor=green,plainpages=false,pdfpagelabels]{hyperref}
\usepackage[pdftex]{graphicx}
\usepackage{epstopdf}
\usepackage[normalem]{ulem}
\usepackage{newlfont}
\usepackage{amsfonts}
\usepackage{epsfig}
\usepackage{mathrsfs}
\usepackage[thinc]{esdiff}
\usepackage{caption}
\usepackage{subcaption}
\newcommand\scalemath[2]{\scalebox{#1}{\mbox{\ensuremath{\displaystyle #2}}}}
\usepackage{times}
\def\tr{\operatorname{tr}}

\DeclareOldFontCommand{\rm}{\normalfont\rmfamily}{\mathrm}

\DeclareMathOperator*{\argmin}{Arg\,min}

\begin{document}

\title {Fundamental speed limits on entanglement dynamics of bipartite quantum systems}
\author{Vivek Pandey}\email{vivekpandey3881@gmail.com}

\affiliation{Harish-Chandra Research Institute, A CI of Homi Bhabha National
Institute, Chhatnag Road, Jhunsi, Prayagraj 211019, India
}
\author{Swapnil Bhowmick}\email{swapnilbhowmick@hri.res.in}
\affiliation{Harish-Chandra Research Institute,  A CI of Homi Bhabha National
Institute, Chhatnag Road, Jhunsi, Prayagraj 211019, India
}
\author{Brij Mohan}
\email{brijhcu@gmail.com}
\affiliation{Department of Physical Sciences, Indian Institute of Science Education and Research (IISER), Mohali, Punjab 140306, India}

\author{Sohail}\email{sohail.sohail@ttu.edu}
\affiliation{Texas Tech University, Lubbock, TX 79409, United States}
\affiliation{Harish-Chandra Research Institute,  A CI of Homi Bhabha National Institute, Chhatnag Road, Jhunsi, Prayagraj 211019, India
}
\author{Ujjwal Sen}\email{ujjwal@hri.res.in}
\affiliation{Harish-Chandra Research Institute,  A CI of Homi Bhabha National
Institute, Chhatnag Road, Jhunsi, Prayagraj 211019, India
}
\begin{abstract}
The speed limits on entanglement are defined as the maximal rate at which entanglement can be generated or degraded in a physical process. We derive the speed limits on entanglement, using the relative entropy of entanglement and trace-distance entanglement, for unitary as well as for arbitrary quantum dynamics, where we assume that the dynamics of the closest separable state can be approximately described by the closest separable dynamics of the actual dynamics of the system. For unitary dynamics of isolated bipartite systems which are described by pure states, the rate of entanglement production is bounded by the product of fluctuations of the system's driving Hamiltonian and the surprisal operator, with an additional term reflecting the time-dependent nature of the closest separable state. Removing restrictions on the purity of the input and on the unitarity of the evolution, the two terms in the bound get suitably altered.
 Furthermore, we find a lower bound on the time required to generate or degrade a certain amount of entanglement by arbitrary quantum dynamics. We demonstrate the tightness of our speed limits on entanglement by considering quantum processes of practical interest.
\end{abstract}
\maketitle

\section{Introduction}

Quantum speed limits (QSLs) are the fundamental limitations imposed by quantum mechanics on the rate at which quantum mechanical systems and their observables evolve in a physical process. Initially, QSLs were formulated for unitary dynamics of pure states~\cite{Mandelstam1945, Margolus1998, Anandan1990, Pati2023, Dimpi20222}. Later it was generalized for unitary dynamics of mixed states~\cite{UHLMANN1992}, entangling dynamics~\cite{Giovannetti_2004,Canseco_J_2022,Krisnanda_2022}, dissipative quantum dynamics~\cite{ Taddei2013, del_Campo2013, Deffner_2013, Peris2016, Naoto2021, Funo_2019,Saito2023, Peris2016, Brody_2019,Campaioli2013,Kiselev2022}, and arbitrary dynamics~\cite{Dimpi2022}. The QSLs are essential for the theoretical understanding of quantum dynamics and have substantial practical relevance in developing quantum technologies. QSLs have several applications in various quantum technologies such quantum computing~\cite{AGN12,Aifer2022}, quantum metrology~\cite{Campbell2018,Beau2017}, optimal control theory~\cite{Caneva2009,deffner_2017}, quantum communication~\cite{Murphy2010}, quantum batteries~\cite{Mohan_Pati_2021,Mohan2022,Modi2017} and quantum refrigerator~\cite{Mukhopadhyay2018}.

 Entanglement is one of the remarkable features of quantum theory with no counterpart in a classical theory~\cite{EPR1935, Bell1964}. It is a feature of composite systems which implies that there exists a set of global states such that any element of this set can not be written as a mixture of products of states of the individual subsystems. In 1964 John Bell derived an inequality for composite systems assuming that the physical world is described by a local hidden variable model and showed that certain entangled states can violate this inequality~\cite{Bell1964}. From this perspective, entanglement can be seen as a characteristic feature of quantum theory and is impossible to generically simulate in any classical formalism. Entanglement is a key resource for several information processing tasks such as quantum teleportation~\cite{Bennett1993a}, quantum dense coding~\cite{Wiesner1992}, quantum cryptography~\cite{Ekert1991}, quantum communication~\cite{Wiesner1992}, quantum computation~\cite{Jozsa1997,Jozsa2003}, quantum random number generators~\cite{Colbeck2012}, quantum metrology~\cite{Jonathon2008}, etc. Therefore, understanding the dynamics of entanglement, we can uncover new ways to generate and manipulate it for a variety of applications. Additionally, studying the dynamics of entanglement helps us gain insight into the behavior of quantum systems. There has been a significant amount of interest in studying entanglement dynamics in various physical systems, including many-body and optomechanical systems~\cite{mbed,Elsayed_2018,oped}.

Quantum speed limits on the time evolution of quantum systems were traditionally derived based on the distinguishability of the two quantum states: the initial and final states.
 Recently, some other formulations of QSLs have been studied, which focused on the rate of change of different quantities, such as the expectation value of observables~\cite{Mohan2022,Gong_2022,hamazaki2023quantum, Pintos2022, Hamazaki2022}, auto-correlation function~\cite{Niklas2023, Carabba_2022,Hasegawa2023}, quantumness~\cite{Jing_2016}, quantum coherence~\cite{Brij2022}, entanglement~\cite{Bera2013, Rudnicki20201, vivek2022, Divyansh2022}, quantum resources~\cite{Campaioli_2022}, information measures~\cite{Brij2022, Diego2022, Pires2021}, and correlations~\cite{vivek2022,Paulson2022}.

 Entanglement is an invaluable resource in quantum information and computation. However, it is a seemingly fragile quantum resource and can easily be degraded by undesirable but practically unavoidable interactions with the environment, such as decoherence and dissipation. An area of ongoing research is to understand and control the dynamics of entanglement between two subsystems by tuning their interaction in the presence of noise~\cite{Shaham_2015,Nosrati_2020,Laynon2013}. Similarly, there is a need for further work to speed up the generation of entanglement in various quantum systems that are commonly used in quantum technologies using quantum control methods. Consequently, the study of quantum speed limits on entanglement generation, manipulation, and degradation is of critical significance.

The speed limits on entanglement dynamics answer the fundamental question, ``how fast can  entanglement be generated or degraded in a physical process?" The speed limits on entanglement dynamics are typically obtained as upper bounds on the rate of change of entanglement of a quantum system in a physical process. Thus, they also provide lower bounds on the time required to change the entanglement of a quantum system by a certain amount. Traditionally, the speed limits on entanglement dynamics were studied for unitary dynamics of quantum systems using suitable entanglement measures such as entanglement entropy~\cite{Divyansh2022,Hamilton2023,vivek2022}, concurrence~\cite{vivek2022}, and geometric measure of entanglement~\cite{Bera2013, Rudnicki20201}. Recently, speed limits on entanglement have been generalized to arbitrary dynamics using entanglement measures such as negativity~\cite{vivek2022}. Negativity, however, cannot quantify the entanglement of bound-entangled states\cite{Horodicki1998}. It is, therefore, interesting to derive the speed limits on entanglement using a suitable entanglement measure that does not have such shortcomings. In Ref.~\cite{Campaioli_2022}, a resource speed limit is introduced, which describes how fast quantum resources can be generated or degraded by physical processes. The resource speed limit is derived from a relative entropy-based measure, which also applies to entanglement. In general, when the state of the given system changes in time, the free state, which minimizes the relative entropy, also changes at each instant of time. In Ref.~\cite{Campaioli_2022}, while deriving the resource speed limit, the authors considered the free state as fixed (stationary) for simplicity. Therefore, the bounds presented in Ref.~~\cite{Campaioli_2022}, while very important and general, can potentially be loose.

In the context of the relative entropy of entanglement\cite{Vedral1997}, the free states are separable states. We focus on the dynamics of the closest separable state, which minimizes the relative entropy of entanglement at each instant of time when the state of a given quantum system evolves. Any map which describes the dynamics of the closest separable state is potentially one that maps separable states to separable states. In other words, the dynamics of the closest separable state is considered to be an entanglement non-generating map for separable initial states. This can, in principle, be a LOCC (local operations and classical communication) map or a separable map. In this paper, we first assume that the dynamics of the closest separable state can be modeled by the nearest separable map of the given dynamics, which governs the evolution of the given bipartite system. Further in this paper, we derive speed limits on the entanglement of a bipartite system evolving under arbitrary dynamical processes.
To obtain the speed limits on entanglement, we use the relative entropy of entanglement~\cite{Vedral1997} and the trace distance of entanglement~\cite{Eisert_2003} as entanglement measures. We have estimated speed limits on entanglement for some quantum processes of practical interest, such as unitary dynamics generated by a non-local Hamiltonian and pure dephasing dynamics.

The structure of this paper is as follows. In Section~\ref{preliminaries}, we present the preliminaries and background necessary to derive the main results of this paper. In Section~\ref{closest separable state}, we provide a method to determine the closest separable dynamics to a given dynamics. In Section~\ref{quantum speed limit}, we derive the speed limits on entanglement for unitary and arbitrary completely positive trace-preserving (CPTP) dynamics using the relative entropy of entanglement and the trace distance of entanglement.  In Section~\ref{examples}, we discuss some examples to show the tightness of our speed limit bounds. Finally, in Section~\ref{conclusion}, we summarise our findings.

\section{ Definitions and  Preliminaries}\label{preliminaries}
 In this section, we discuss the preliminaries necessary to arrive at the main result of the article. For completeness, we have provided additional preliminary details in Appendix~\ref{prelims}. 

 Let $\cal{H}$ denote a separable Hilbert space with ${\rm dim}\left(\cal{H}\right)$ denoting the dimension of $\cal{H}$. Let $\cal{L}(\cal{H})$ denotes the set of all linear operators acting on $\cal{H}$, with $\cal{I}_{\cal{H}}$ denoting the identity operator. The subset of $\cal{L}(\cal{H})$, containing all positive semi-definite operators is denoted by $\cal{L}(\cal{H})_{+}$. Let ${\cal{L}}'(\cal{H})_{+}$ denote the subset of $\cal{L}(\cal{H})$ containing all positive semi-definite operators acting on $\cal{H}$. The state of a quantum system, denoted by $\rho$, is an element of ${\cal{L}'(\cal{H}})_{+}$ with unit trace and is called a density operator. We denote with ${\cal{S}(\cal{H})}$, the set of density operators acting on the Hilbert space $\cal{H}$. Let $\cal{H}_{A}$ denote the Hilbert space associated with a quantum system $A$. States of the system $A$ are represented by density operators $\rho_{A}\in{\cal{S}(\cal{H}_{A})}$. For a composite system $AB$, the combined Hilbert space is the tensor product $\cal{H}_{AB}:=\cal{H}_{A}\otimes\cal{H}_{B}$ of the individual Hilbert spaces. A state of the composite system $AB$ is represented by a density operator $\rho_{AB}\in \cal{S}(\cal{H}_{AB})$, and the reduced density operator of the subsystem $A$ (or $B$) is obtained by applying the partial trace operation $\tr_{B}$ (or $\tr_{A}$) over the combined density operator $\rho_{AB}$, i.e., $\rho_{A}=\tr_{B}(\rho_{AB})$, and similarly for $B$. A quantum system is said to be in a pure state when there is no uncertainty in the knowledge of its state. The density operator representing a pure state $\Psi\in\mathcal{S}(\mathcal{H})$ is a rank-one projection operator i.e. $\Psi = \op{\Psi}$~(an idempotent matrix, i.e., $\Psi^{2}=\Psi$), where $\ket{\Psi}\in\mathcal{H}$. Otherwise, $\rho$ is in a mixed state and can be written as a convex mixture of pure states $\Psi_{i}$ with mixing probabilities $p_{i}$, i.e., $\rho  =\sum_{i}p_{i}\Psi_{i}$. In this paper, we will denote pure states with $\Psi$.

{\it Separable states, separable operations and entanglement}.-- A bipartite quantum state $\sigma_{AB}$ is called separable~(or completely disentangled) if it can be written or approximated in the following form~\cite{Werner1989}:
\begin{equation}
    \sigma_{AB} = \sum_{i}p_{i}\rho^i_A \otimes \rho^i_B ,
\end{equation}
for some probability distribution $\{p_{i}\}_i$ and density operators $\rho^i_A$ and $\rho^i_B$. When subsystems $A$ and $B$ are finite-dimensional, we can choose $\rho^i_A$ and $\rho^i_B$ to be pure states. We let $\operatorname{SEP}(A;B)$ denote the set of separable states defined on $\mathcal{H}_{AB}$. An operation is called ``separable operation" if its Kraus representation is given by:
\begin{equation}
    \Lambda_S (\rho)= \sum_{i} A_i \otimes B_i \rho A_i^\dag \otimes B^\dag_i ,
\end{equation}
which satisfy $\sum_{i}A_{i}A^{\dag}_{i}\otimes B_{i}B^{\dag}_{i}=\cal{I}\otimes\cal{I}$.

If $\sigma_{AB}$ is not separable, then it is ``entangled".
 Entanglement works as a valuable resource for many quantum information processing tasks such as quantum teleportation, quantum metrology, quantum communication, quantum sensing, quantum random number generators, etc. So, it is desirable to have a measure that quantifies the amount of entanglement present in a state. There are several measures of entanglement which have been proposed namely entanglement entropy~\cite{Bennett1996}, concurrence~\cite{Wootters2001,Wootters1998}, negativity~\cite{Vidal2002}, relative entropy of entanglement~\cite{Vedral1997}, trace distance of entanglement~\cite{Eisert_2003}, etc. In this paper, we will consider the relative entropy of entanglement and trace distance of entanglement as measures of
entanglement.

{\it Relative entropy of entanglement}.--  The relative entropy of entanglement~(REE) is an entanglement measure to quantify the amount of entanglement present in an arbitrary bipartite quantum state. For an arbitrary bipartite quantum state $\rho_{AB}\in {\cal{S}}({\cal{H}}_{AB})$, its relative entropy of entanglement  is defined as~\cite{Vedral1997}
\begin{equation}
    E(\rho_{AB}):= \min_{\sigma \in \operatorname{SEP}(A;B) } D\left(\rho_{AB}\Vert\sigma_{AB}\right),\label{equ:Relative_entropy_of_entanglement}
\end{equation}
where each of $\dim(\cal{H}_A)$ and $\dim(\cal{H}_B)$ can be either finite or infinite. For a given quantum state $\rho_{AB}$, let us assume that minimization of the above equation is achieved for the separable state $\sigma^{*}_{AB}\in {\cal{S}}({\cal{H}}_{AB})$, i.e., $  E(\rho_{AB})=D\left(\rho_{AB}\Vert\sigma^{*}_{AB}\right)$, which depends on the given state $\rho_{AB}$, i.e.,  $\sigma^{*}_{AB}=\sigma^{*}_{AB}(\rho_{AB})$. A separable state that realizes the minimum of Eq.~\eqref{equ:Relative_entropy_of_entanglement}  is called a closest separable state~(CSS).

{\it Trace Distance of entanglement.--} Trace distance can also be used as a distance measure to define an entanglement monotone. It is a distance-based measure defined on the intuition that if a given state is geometrically close to the set of separable states it has less entanglement. For an arbitrary bipartite quantum state $\rho_{AB}\in {\cal{S}}({\cal{H}}_{AB})$, its trace distance of entanglement  is defined as~\cite{Eisert_2003}:
\begin{equation}
    \mathcal{D}_{tr}(\rho_{AB}) = \min_{\sigma \in \operatorname{SEP}(A;B) } \tr \left|\rho_{AB} - \sigma_{AB}\right|.
\end{equation}

We have assumed that the dynamics of the closest separable state (CSS) can be approximated by the closest separable dynamics of a given dynamics. In the next section, we introduce a method to determine the closest separable dynamics of a given arbitrary completely positive trace-preserving (CPTP) dynamics. 
We first find the Choi dual of the given arbitrary CPTP map and then we use relative entropy to find its CSS. Using the Choi-Jmaio{\l}kowski isomorphism, we then find the CPTP map corresponding to the CSS. We call this CPTP map as the closest separable map for the given CPTP map. Then we study the dynamics of the CSS of a given state under the closest separable map obtained in this way.

\section{Dynamics of closest separable state}\label{closest separable state}
We start by finding the closest separable channel to a given channel by using Choi-Jamio{\l}kowski isomorphism. For a given linear CP map $\mathcal{E}:{\cal{L}}'({\cal{H}}_{AB})_{+}\rightarrow {\cal{L}}'({\cal{H}}_{AB})_{+}$, the normalized CJ dual state is given by
\begin{equation}
    \Phi_{AA'BB'}=\frac{1}{d^2}({\rm id}_{A'B'} \otimes \mathcal{E}_{AB})(P_{AA'}\otimes P_{BB'}),
\end{equation}
where $P_{AA'}$ and $P_{BB'}$ are maximally entangled states of $AA'$ and $BB'$ respectively, normalized to $\mathrm{min}\{\mathrm{dim(\mathcal{H}_{A})},\mathrm{dim(\mathcal{H}_{A'})}\}$ and $\mathrm{min}\{\mathrm{dim(\mathcal{H}_{B})},\mathrm{dim(\mathcal{H}_{B'})}\}$. We now find the closest separable state of this Choi operator using relative entropy as the distance measure, which is given as  

\begin{equation}
    \Tilde{\Phi}= \argmin\{S(\Phi\Vert\chi) : {\chi \in \operatorname{SEP}(AA';BB')}\}.
\end{equation}
We define the closest separable map of given map $\mathcal{E}$ as the Choi dual of $\Tilde{\Phi}$, given by
\begin{equation}
    \mathcal{E}^* (\rho_{AB})=d^2 \tr_{A'B'}(\Tilde{\Phi}_{AA'BB'}\rho^T_{A'B'}\otimes\mathcal{I}_{AB}),
\end{equation}\\
where $\mathcal{E}^*$ is the closest separable map of $\mathcal{E}$, its action on state $\rho_{AB}$ is given by the above equation, $\mathcal{I}_{AB}$ is identity operator on subsystem $AB$ and we have taken the dimension of all the systems to be $d$.

As an example, now we will determine the closed separable map of the unitary map. Let us consider a unitary map $\mathcal{E}$  represented by $U_{AB}= {\rm e}^{-iH_{AB}t}$, where $H_{AB}$ is the Hamiltonian of $AB$, . In this case, $\Phi$ is pure, and an analytic expression of the closest separable state is known \cite{Vedral}. For a given pure state in the Schmidt basis $\{\ket{\psi_{i}}\ket{\phi_{i}}\}$, and Schmidt coefficients $\sqrt{p_{i}}$, the density matrix is $\rho= \sum_{i,j} \sqrt{p_i p_j}\ketbra{\psi_i \phi_i}{\psi_j \phi_j}$, and the closest separable state is
\begin{equation}
    \rho^* = \sum_i p_i \ketbra{\psi_i \phi_i}{\psi_i \phi_i}. 
\end{equation}

Let us choose the bases on the Hilbert spaces involved as follows:
\begin{align}
    \mathcal{H}_{A}&=[\{\ket{\alpha_i}\}], \mathcal{H}_{A'}=[\{\ket{\alpha'_j}\}],\nonumber \\  \mathcal{H}_{B}&=[\{\ket{\beta_m}\}],
    \mathcal{H}_{B'}=[\{\ket{\beta'_n}\}]. 
\end{align}
Let the eigenstates of $H_{AB}$ be given by $\{\ket{e_n}_{AB}\}$ with the corresponding eigenvalues $\{E_n\}$. We also define the inner products between the energy eigenbasis and the product basis corresponding to the choice in the above equation as:
\begin{equation}
    V^k_{ij}=\braket{\alpha_i \beta_j}{e_k}_{AB}.
\end{equation}
In terms of the above definitions, the Choi dual of $U_{AB}$ is a pure state in $\mathcal{H}_{AB}\otimes\mathcal{H}_{A'B'}$ and given by:
\begin{equation}
    \Phi_{ABA'B'}=\sum_{i,j,m,n} R_{mi,nj}\ket{\alpha_m \alpha'_i}_{AA'} \ket{\beta_n \beta'_j}_{BB'}, \label{choiunitary}
\end{equation}
where 
\begin{equation}
    R_{mi,nj}= \sum_k {V^k_{ij}}^* V^k_{mn} e^{-iE_{k}t}.
\end{equation}

In order to find the closest separable state of $\Phi_{ABA'B'}$, we need to find the Schmidt decomposition of the state in Eq. \eqref{choiunitary} with respect to the bipartitions $AA':BB'$. We need to choose a particular Hamiltonian for that. Then, the closest separable state is known in terms of the Schmidt basis of $\Phi_{ABA'B'}$. The Choi dual of the closest separable state of $\Phi_{ABA'B'}$ will give the closest separable operation.

In the case of mixed states or non-unitary evolution, finding the analytical expression for the closest separable operation of a given map is, in general, difficult, but one can use the results of \cite{Fried2011} to obtain an entangling operation on $\rho_{AB}$ corresponding to a given closest separable evolution of the closest separable state (CSS) $\sigma^*_{AB}$.

\section{ Speed Limits on Entanglement}\label{quantum speed limit}
 In general, quantum speed limits are the fundamental dynamical constraints imposed by quantum physics on the rate at which states and observables of quantum systems change over time. They provide the lower bound on the time required for changes in the distinguishability of states or changes in the expectation value of observables. The quantum speed limits for states and observables have been extensively studied for both unitary and non-unitary dynamics~\cite{Taddei2013, del_Campo2013, Deffner_2013, Peris2016, Pintos2022}. However, while quantum states and observables evolve over time in physical processes, entanglement may not change or may change at a slower rate. Therefore, conventional speed limits are not applicable to changes in entanglement. In this section, we present distinct quantum speed limits on entanglement manipulations in physical processes.

 Entanglement is a non-classical correlation present between bipartite or multipartite systems, which acts as a resource for several quantum processing tasks such as quantum teleportation~\cite{Ekert1991}, quantum key distribution~\cite{Das2021}, quantum communication~\cite{Bennett1993a}, quantum computation~\cite{Jozsa1997, Jozsa2003}, etc. 
 Generating the desired amount of entanglement on various experimental platforms requires high control, especially in the presence of unwanted interactions and noise. Another challenge in quantum communication and quantum computing is maintaining entanglement over long distances and long periods in the presence of noise~\cite{Yu2006, Gatto2019, Sakuldee2023}. Subsystems become uncorrelated due to the presence of external noises, such as decoherence and dissipation. Moreover, the current interest lies in the fast generation of entangled states~\cite{Sorelli2019, PaweL2023}.
 To this end, a natural question arises as to ``how quickly can entanglement be generated or destroyed by some physical process?'' To answer this question, quantum speed limits on entanglement are introduced. The quantum speed limits on entanglement are defined as the maximal rate of change of entanglement, or upper bounds thereof, and provide lower bounds on the time required for a certain change in entanglement. 

 The speed limits on entanglement have been investigated earlier for unitary dynamics of bipartite quantum systems using suitable entanglement measures such as entanglement entropy~\cite{Divyansh2022, Hamilton2023}, concurrence~\cite{vivek2022} and geometric measure of entanglement~\cite{Bera2013, Rudnicki20201}. In Ref.~\cite{vivek2022}, the speed limits on entanglement have been generalized to arbitrary dynamics using entanglement measures such as negativity. The speed limits on entanglement obtained in Ref.~\cite{vivek2022} can not be used to obtain the speed limits on entanglement for bound entangled states because negativity is unable to quantify the entanglement of bound entangled states. In Ref.~\cite{Campaioli_2022}, the speed limit on entanglement has been obtained using the relative entropy of entanglement (REE). Note that for any dynamical system with associated Hilbert space $\cal{H}$, the state of the system depends on time. So, in general, the closest free state will also be time-dependent. The speed limit bounds in \cite{Campaioli_2022} are obtained on the assumption that during the entire evolution period, the closest separable state is stationary. Therefore, the speed limit bounds in \cite{Campaioli_2022} may not correctly identify the quantities on which the speed limit on entanglement depends. Moreover, in previous works, bounds on the entangling rate have been obtained for the unitary dynamics of pure bipartite systems~\cite{Bravyi2007, Gong2022}. In this work, we also generalize the bound on the entangling rate beyond pure states, as well as unitary dynamics.

In this section, we have used the relative entropy of entanglement (REE) to derive speed limits on entanglement. Our analysis assumes that the evolution of the closest separable state is governed by the closest separable map, which is the map that is closest to the actual map governing the state's evolution. We have applied this assumption to obtain the speed limits on entanglement for both unitary and arbitrary completely positive trace-preserving (CPTP) dynamics.

\subsection{Speed limit on the relative entropy of entanglement for arbitrary CPTP dynamics}

We consider the bipartite quantum system $AB$ with associated Hilbert space ${\cal{H}}_{AB}$, where ${\rm dim}({\cal{H}}_{AB})$ is finite. The dynamics of the bipartite system $AB$ in a time interval $I=[t_{0},t_{1}]$, is governed by a CPTP map $\cal{N}_{\{{t_{0},t}\}}$, where $t\in I$. Let $\rho_{t_0}$ be the initial state of the bipartite system and $\sigma^{*}_{t_0}$ be its closest separable state. The state of the system $\rho_{t}$ at any time $t$ is given by $\rho_{t}=\cal{N}_{\{{t_{0},t}\}}(\rho_{t_{0}})$. Let $\cal{N}^{*}_{\{{t_{0},t}\}}$ be the closest separable map of $\cal{N}_{\{{t_{0},t}\}}$. Assuming that the relative entropy $ D\left(\cal{N}_{\{{t_{0},t}\}}(\rho_0)\Vert\cal{N}^{*}_{\{{t_{0},t}\}}(\sigma^{*}_{0})\right)$ is well defined (i.e., $\operatorname{supp}(\rho_t)\subseteq \operatorname{supp}(\sigma_t)$) for all $t\in I$, the relative entropy of entanglement of the bipartite system at time $t$ is given by

\begin{align}
    D\left(\cal{N}_{\{{t_{0},t}\}}(\rho_0)\Vert\cal{N}^{*}_{\{{t_{0},t}\}}(\sigma^{*}_{0})\right) &=D\left(\rho_{t}\Vert\sigma_{t}\right)\nonumber\\ &=\tr(\rho_{t}\log{\rho_{t}}-\rho_{t}\log{\sigma_{t}})\nonumber\\
    &=-S(AB)_{\rho_t}-f(t),
\end{align}
where $\sigma_{t}=\cal{N}^{*}_{\{{t_{0},t}\}}(\sigma^{*}_{0})$, $S(AB)_{\rho_t}=-\tr(\rho_{t}\log{\rho_{t}})$, and $f(t)=:\tr(\rho_t\log \sigma_t)$. We further assume that derivative of $D\left(\rho_{t}\Vert\sigma_{t}\right)$ is well defined in $t\in I$. After differentiating both sides of the above equation with respect to time $t$, we obtain
\begin{align} \label{entropyrate}
    \frac{\rm d}{{\rm d}t} D\left(\rho_{t}\Vert\sigma_{t}\right)&=-\left(\frac{\rm d}{{\rm d}t}S(AB)_{\rho_t}+\frac{\rm d}{{\rm d}t}f(t)\right)\nonumber\\
    &=-\Gamma(t)-\frac{\rm d}{{\rm d}t}f(t),
\end{align}
where $\Gamma(t)=\frac{\rm d}{{\rm d}t}\left(S(AB)_{\rho_t}\right)=-\tr\left(\dot{\rho}_t{\Pi}_{\rho_t}\log \rho_t\right)$ is the rate of change of entropy~\cite{Das2018}, ${\Pi}_{\rho_t}$ is projection onto the support of $\rho_t$ and we have assumed that $\dot{\rho}_t$ is well defined. The derivative of $f(t)$ can be written as~(see \eqref{derivative_of_f}) 
\begin{equation}
    \frac{{\rm d}}{{\rm d}t} f(t)= \tr (\Dot{\rho_t}{\Pi}_{\sigma_t}\log \sigma_t) + \tr(\rho_t {\Pi}_{\sigma_t}G_{\sigma_t}(\Dot{\sigma}_t)).
\end{equation}
In Appendix \ref{derivation_of_D(f(t))}, we have provided a detailed calculation of $\frac{{\rm d}}{{\rm d}t} f(t)$.
Let ${\cal{F}}_E(t):=D\left(\rho_{t}\Vert\sigma_{t}\right)$, and then the time derivative of ${\cal{F}}_E(t)$ can be written as 
\begin{align} \label{entrate2}
      \frac{\rm d}{{\rm d}t} {\cal{F}}_{E}(t)& = \scalemath{0.95}{-\Gamma(t)-\tr (\Dot{\rho_t}{\Pi}_{\sigma_t}\log \sigma_t) - \tr(\rho_t {\Pi}_{\sigma_t}G_{\sigma_t}(\Dot{\sigma}_t))}.
\end{align}
By taking the absolute value of both sides of the above equation and applying the triangle and Cauchy-Schwarz inequalities, we can obtain
\begin{align}
    \left|\frac{\rm d}{{\rm d}t} {\cal{F}}_E(t)\right|
    &\leq \norm{\mathcal{L}_{t}({\rho_t})}_{\rm 2} \norm{{\Pi}_{\rho_t}\log \rho_t-{\Pi}_{\sigma_t}\log \sigma_t}_{\rm 2}\nonumber\\
    &\hspace{0.35cm}+\left|\tr(\rho_t {\Pi}_{\sigma_t}G_{\sigma_t}(\Dot{\sigma}_t))\right|.
\end{align}
The operators $\log \rho_t$ and $\log \sigma_t$ are defined on the supports of $\rho_t$ and $\sigma_t$, respectively. Then we have
\begin{equation}
    \scalemath{0.99} { \left|\frac{\rm d}{{\rm d}t} {\cal{F}}_{E}(t)\right|\leq \norm{\mathcal{L}_{t}({\rho_t})}_{\rm 2} \norm{\log \rho_t-\log \sigma_t}_{\rm 2}+\left|\tr(\rho_t G_{\sigma_t}(\Dot{\sigma}_t))\right|}\nonumber. \label{equ:relative_entropy_of_entanglement_rate}
\end{equation}
To interpret the above inequality, first note that $\norm{\mathcal{L}_{t}({\rho_t})}_{\rm 2}$ is the evolution speed of the quantum system~\cite{Campaioli_2019}. Thus, the entanglement rate is upper bounded by the product of the speed of the quantum evolution and the 2-norm of difference of two surprisal operators, viz.~$\left(\norm{\ln \rho_t-\ln \sigma_t}_{\rm 2}\right)$. There is also a time-dependent additional term ($\left|\tr(\rho_t G_{\sigma_t}(\Dot{\sigma}_t))\right|$)  due to the evolution of the closest separable state (CSS), which will vanish if the operator $\sigma_t$ is time-independent. The above inequality provides an upper bound on the maximal rate at which any physical process can generate or deplete entanglement in a bipartite system. In other words, the rate of change of entanglement cannot exceed the limit imposed by this inequality. By considering this bound, we can gain insights into the fundamental limits of entanglement generation and depletion in quantum systems. 

Let us integrate the above inequality with respect to time $t$ in the range $[0,T]$ and we then obtain the following bound, 
\begin{align}
    \int_{0}^{T}  \left|\frac{\rm d}{{\rm d}t} {\cal{F}}_{E}(t)\right|{{\rm d}t}& \leq \int_{0}^{T}  \mathbbm{k}(t){\rm d}t, \label{equ:integral_over_t}
\end{align}
where $\mathbbm{k}(t):= \norm{\mathcal{L}_{t}({\rho_t})}_{\rm 2} \norm{\log \rho_t-\log \sigma_t}_{\rm 2} +\left|\tr(\rho_t G_{\sigma_t}(\Dot{\sigma}_t)\left.\right)\right|$.
From the above inequality, we obtain
\begin{equation}
     T \ge T^{\mathbbm{k}}_{\rm ESL} := \frac{|{\cal{F}}_{E}(T)-{\cal{F}}_{E}(0)|}{\Lambda^E(T)},\label{equ:speeed_limit_entanglemen}
\end{equation}
where $\Lambda^E(T):={\frac{1}{T} \int_{0}^{T} \mathbbm{k}(t) {\rm d}t}$.

The bound described above provides the minimum time scale required to bring a certain change in the entanglement of a bipartite system using any physical process. 

{\it Minimal time required for the generation of entanglement}.--

Let us assume that ${\cal{S}}_{\{t_0,t\}}$ is the actual map which governs the dynamics of the closest separable state. Here, we approximated the actual map ${\cal{S}}_{\{t_0,t\}}$  by the closest separable map ${\cal{N}}^{*}_{\{t_0,t\}}$ of the given dynamics. It is straightforward to check that the following inequality holds for all $t\in I$

 \begin{equation}
 E(\rho_t) \leq {\cal{F}}_{E}(t), \label{equ:relative_entropy_of_entanglemeny_inequality}
 \end{equation}
 where ${\cal{F}}_{E}(t)=D\left({\cal{N}_{\{{t_{0},t}\}}}(\rho)\Vert {\cal{N}^*_{\{{t_{0},t}\}}}(\sigma^*_{0})\right)$, $E(\rho_t)=D\left({\cal{N}_{\{{t_{0},t}\}}}(\rho)\Vert{\cal{S}}_{\{t_0,t\}}(\sigma^*_{0})\right) $, and $E(\rho_0)={\cal{F}}_{E}(0) $.
  If the initial state of the system is separable or product, then
using inequality~\eqref{equ:relative_entropy_of_entanglemeny_inequality} and bound~\ref{equ:speeed_limit_entanglemen}, we obtain a lower bound on the time required to generate a certain amount of entanglement, which is given by

\begin{align}
    T \ge T^{g}_{\rm ESL} = \frac{{\cal{F}}_{E}(T)}{\Lambda^E(T)}\geq \frac{E(\rho_T)}{\Lambda^E(T)}.
 \label{speeed_limit_entanglement_production}
\end{align}

The bound described above provides the minimum time required to generate a given amount of entanglement in a bipartite system from a separable or product state using any physical process. Moreover, the bound serves as a fundamental constraint on the entanglement dynamics in such systems, ensuring that any physical process cannot generate entanglement faster than the bound allows.

{\it Minimal time required for the degradation of entanglement}.--
If the initial state of the system is entangled and the final state is separable, then
using inequality~\eqref{equ:relative_entropy_of_entanglemeny_inequality} and bound~\ref{equ:speeed_limit_entanglemen}, we obtain a lower bound on the time required to completely deplete the given amount of entanglement, which is given by
\begin{align}
    T \ge T^{dg}_{\rm ESL} := \frac{{\cal{F}}_{E}(0)}{\Lambda^E(T)}\geq \frac{E(\rho_0)}{\Lambda^E(T)}.
 \label{speeed_limit_entanglement_depletation}
\end{align}
The bound described above provides a tool for estimating the minimum survival time of entanglement in a bipartite system that is evolving under the influence of local noise. Additionally, the lower bound can be used to estimate the time required for entanglement breaking by a dynamical map, a concept recently discussed in Ref~\cite{Lukasz2023}. As such, the bound offers valuable insight into the dynamics of entanglement in bipartite systems subjected to various forms of noise and decoherence.

Now, we consider the special case when the bipartite system $AB$ is an isolated system.
\subsection{Speed limit on the relative entropy of entanglement manipulation for unitary dynamics}
Let us consider a bipartite quantum system whose dynamics is governed by a unitary map. We can use Eq.~\eqref{entrate2} to calculate the entanglement rate in such a scenario. Since the von Neumann entropy is  invariant under unitary dynamics, $\Gamma(t)$ will vanish,  and the entanglement growth rate for unitary dynamics is given by
\begin{equation}
      \frac{\rm d}{{\rm d}t} {\cal{F}}_{E}(t)=-\tr (\Dot{\rho_t}{\Pi}_{\sigma_t}\log \sigma_t) - \tr(\rho_t {\Pi}_{\sigma_t}G_{\sigma_t}(\Dot{\sigma}_t))
\end{equation}
Taking the absolute value on both sides of the above equation and then applying the triangle inequality, we get
\begin{equation}
     \left|\frac{\rm d}{{\rm d}t} {\cal{F}}_{E}(t)\right|\leq \left|\tr(\Dot{\rho_{t}}\log{\sigma_{t}})\right|+\left|\tr(\rho_{t}G_{\sigma_t}(\Dot{\sigma}_t))\right|.\label{equ:equ:unitary_rate_1}
\end{equation}
For unitary dynamics, the evolution of the state is described by the Liouville-von Neumann equation,
\begin{equation}
     \Dot{\rho_{t}} = -i[H,\rho_{t}],\label{equ:Liouvill_von_Neumann_equation}
\end{equation}
where $H$ is the driving Hamiltonian of the quantum system, and we have taken $\hbar = 1$. Using  Eq.~\eqref{equ:Liouvill_von_Neumann_equation} and  inequality~\eqref{equ:equ:unitary_rate_1}, we obtain
\begin{align}
 \left|\frac{\rm d}{{\rm d}t} {\cal{F}}_{E}(t)\right|&\leq \left|\tr([H,\rho_{t}]\log{\sigma_{t}})\right|+\left|\tr(\rho_{t}G_{\sigma_t}(\Dot{\sigma}_t))\right|\nonumber\\
    &=|\tr([\log{\sigma_{t}},H]\rho_{t})|+|\tr(\rho_{t}G_{\sigma_t}(\Dot{\sigma}_t))|.\nonumber\\
\end{align}

Using the uncertainty relation~\eqref{equ:uncertainty_relation_for_mixed_states} ( for ${\cal{O}}_{1}=\log \sigma_t$ and ${\cal{O}}_{2}= H $), the above inequality leads to
\begin{align}
    \left|\frac{\rm d}{{\rm d}t} {\cal{F}}_{E}(t)\right|&\leq 2\sqrt{U_{\log{\sigma_{t}}}U_{H}}+ \left|\tr(\rho_{t}G_{\sigma_t}(\Dot{\sigma}_t))\right|.\label{equ:entanglement_rate_for_unitary_dymanics}
\end{align}
If the initial state of the system is pure, denoted by $\Psi_{0}$ , the above bound reduces to the form 
\begin{align}
   \left|\frac{\rm d}{{\rm d}t} {\cal{F}}_{E}(t)\right|&\leq 2\left(\Delta \log{\sigma_{t}}\right) \left(\Delta H\right) + \left|\tr(\Psi_{t}G_{\sigma_t}(\Dot{\sigma}_t))\right|,\label{equ:entanglement_rate_for_unitary_dymanics_pure}
\end{align}
where $\Delta A =\sqrt{\langle A^2 \rangle - \langle A \rangle^2}$ is the variance of operator $A$ and $\langle A \rangle = \tr(\Psi A)$ is the  expectation value function of $A$ on $\Psi$, and $\Psi_t$ is state of the system at time $t$.

We can interpret the above inequality by breaking it down into its constituent parts. $2\Delta H$ represents the speed at which the system's state is transported in the projective Hilbert space \cite{Anandan1990}, while $\Delta \log{\sigma_{t}}$ is the variance of the ``surprisal" operator \cite{Pintos2022}. Thus, the rate of entanglement is upper-bounded by a product of the speed of quantum evolution, the variance of the surprisal operator, along with a time-dependent additional term arising from the evolution of the closest separable state (CSS). This correction term vanishes if $\sigma_t$ is time-independent. The bounds given in \ref{equ:entanglement_rate_for_unitary_dymanics} and \ref{equ:entanglement_rate_for_unitary_dymanics_pure} provide an upper bound on the maximal rate at which any non-local unitary operation can generate or deplete entanglement in a bipartite system. 

Let us integrate the inequality \eqref{equ:entanglement_rate_for_unitary_dymanics} with respect to time $t$, to obtain the following bound: 
\begin{equation}
    \int_{0}^{T}  \left|\frac{\rm d}{{\rm d}t} {\cal{F}}_{E}(t)\right|{{\rm d}t} \leq \int_{0}^{T}  \mathbbm{u}(t){\rm d}t,
\end{equation}
where $\mathbbm{u}(t)= 2\sqrt{U_{\log{\sigma_{t}}}U_{H}}+ \left|\tr(\rho_{t}G_{\sigma_t}(\Dot{\sigma}_t))\right|$.

From the above inequality, we obtain
\begin{equation}
     T \ge T_{\rm ESL} := \frac{|{\cal{F}}_{E}(T)-{\cal{F}}_{E}(0)|}{\Lambda^E_{\mathbbm{u}}(T)},\label{equ:speeed_limit_entanglemen_unitary}
\end{equation} 
where $\Lambda^E_{\mathbbm{u}}(T):={\frac{1}{T} \int_{0}^{T} \mathbbm{u}(t) {\rm d}t}$.

The bound in \ref{equ:speeed_limit_entanglemen_unitary} provides the minimum time scale for certain changes in the entanglement of a bipartite system by any unitary process. From \ref{equ:speeed_limit_entanglemen_unitary}, bounds similar to \ref{speeed_limit_entanglement_production} and \ref{speeed_limit_entanglement_depletation} can be obtained for unitary dynamics which provides a lower bound on the minimum time required to generate a certain amount of entanglement (staring from the separable state) or fully deplete a certain amount of entanglement.

Note that all the calculations up to this point have been done assuming that the support of $\sigma_t$ must be greater than or equal to the support of $\rho_t$ for all $t\in [t_0,t_1]$. However, there can be some CPTP dynamics that violate this assumption, and in that scenario, none of the bounds will be valid. To circumvent this problem, we now take trace distance to define a distance-based entanglement measure to calculate the speed limits on entanglement dynamics.

\subsection{ Speed limit on entanglement using trace distance}
We consider the dynamics of a bipartite system $AB$ in time interval $I \in [t_0,t_1]$. The evolution of the system is described by a CPTP map $\mathcal{N}_{\{t_0,t\}}$, where $t\in I$. Let $\rho_{t_0}$ be the initial state of the system and $\sigma^{*}_{t_0}$ be its closest separable state (CSS) in the trace distance measure.  The dynamics of $\sigma^{*}_{t_0}$  is assumed to be governed by the closest separable map of $\mathcal{N}_{\{t_0,t\}}$. We let  $\mathcal{N}^*_{\{t_0,t\}}$ denote the closest separable map of $\mathcal{N}_{\{t_0,t\}}$. Then, the
trace distance at time $t$ is given by
\begin{align}
   \mathcal{D}_{tr}\left(\mathcal{N}_{\{t_0,t\}}(\rho_{t_0}),\mathcal{N}^*_{\{t_0,t\}}(\sigma^*_{t_0})\right) &=   \mathcal{D}_{tr}(\rho_t,\sigma_t)\nonumber\\
  &= \tr|\rho_t - \sigma_t|.
\end{align}
After differentiating both
the sides of the above equation with respect to time $t$, we obtain
\begin{align}
     \frac{\rm d}{{\rm d}t} \mathcal{D}_{tr}(\rho_t,\sigma_t) =  \frac{\rm d}{{\rm d}t} \tr|\rho_t - \sigma_t|.
\end{align}
Assuming that $\mathcal{D}_{tr}(\rho_t,\sigma_t)$ is a smooth function of $t$, its derivative can be written as 
\begin{align}
    \frac{\rm d}{{\rm d}t} \mathcal{D}_{tr}(\rho_t,\sigma_t) = \lim_{\epsilon\rightarrow 0}
    \frac{\mathcal{D}_{tr}(\rho_{t+\epsilon},\sigma_{t+\epsilon})-\mathcal{D}_{tr}(\rho_t,\sigma_t)}{\epsilon}. \label{equ:trace:derivative}
\end{align}
We can calculate $\mathcal{D}_{tr}(\rho_{t+\epsilon},\sigma_{t+\epsilon})$ explicitly as follows:
\begin{align}
    \mathcal{D}_{tr}(\rho_{t+\epsilon},\sigma_{t+\epsilon}) &= \tr|\rho_{t+\epsilon} - \sigma_{t+\epsilon}|\nonumber\\
    & = \tr|(\rho_t - \sigma_t)+\epsilon(\Dot{\rho_t} - \Dot{\sigma_t})+O(\epsilon^2)|\nonumber\\
    & \leq \tr|\rho_{t} - \sigma_t|+\epsilon \tr|\Dot{\rho_t} - \Dot{\sigma_t}| + \tr(|O(\epsilon^2)|)\nonumber\\
    & = \mathcal{D}_{tr}(\rho_t,\sigma_t) + \epsilon \tr|\Dot{\rho_t} - \Dot{\sigma_t}| + \tr(|O(\epsilon^2)|),\label{equ:trace_inequality}
\end{align}
where the second equality follows from the Taylor expansion of $\rho_{t+\epsilon}$ and $\sigma_{t+\epsilon}$, and in the third step, we have used the triangle inequality. Now, taking absolute values on both the sides of \eqref{equ:trace:derivative} and then using Eq.~\eqref{equ:trace_inequality}, we get~(in limit $\epsilon \rightarrow 0$)
\begin{align}
    \left|\frac{\rm d}{{\rm d}t} \mathcal{D}_{tr}(\rho_t,\sigma_t)\right| \leq 
    \tr|\Dot{\rho_t} - \Dot{\sigma_t}|.
\end{align}
The above inequality gives an upper bound on the rate of entanglement of the bipartite system $AB$ for arbitrary dynamics. By integrating both sides of the above inequality with respect to time $t$, we obtain 
\begin{align}
    \int_{0}^{T}  \left|\frac{\rm d}{{\rm d}t} \mathcal{D}_{tr}(\rho_t,\sigma_t)\right| {\rm d}t\leq \int_{0}^{T}  
    \tr|\Dot{\rho_t} - \Dot{\sigma_t}|{\rm d}t.
\end{align}
From the above inequality, we obtain the bound,
\begin{align}\label{ETD}
    T \ge T^{tr}_{\rm ESL} := \frac{|{\cal{D}}_{tr}(\rho_T,\sigma_T)-{\cal{D}}_{tr}(\rho_0,\sigma_0)|}{\Lambda_{tr}(T)},
\end{align}
where $\Lambda_{tr}(T):=\frac{1}{T}\int_{0}^{T}  
    \tr|\Dot{\rho_t} - \Dot{\sigma_t}|{\rm d}t$.
    
    The upper bound described above represents the minimum timescale required for any physical process to induce a measurable change in the entanglement of a bipartite system. This bound applies to both entanglement generation and entanglement degradation.    

\section{Tightness and attainability of the bounds}\label{examples}
 The tightness and attainability of the bounds obtained from the relative entropy of entanglement depend on the saturation of the inequalities used to derive them, such as uncertainty relations, Cauchy-Schwarz inequality, and triangle inequality. In general, finding the general dynamics that saturate these bounds is quite challenging due to the involvement of the logarithm of operators. Therefore, we demonstrate a few examples to show that these bounds are generally attainable for both unitary and open quantum dynamics. However, the bound~(\ref{ETD}) on the trace distance of entanglement can be easily shown to be attainable when the closest separable state (CSS) does not change over time (see Appendix~\ref{Sat}).

First, we demonstrate the tightness and attainability of bound~(\ref{equ:speeed_limit_entanglemen_unitary}) for unitary dynamics (generated by Heisenberg interaction) for both pure and mixed states.

{\it Unitary dynamics}.--  We consider a two-qubit systems $AB$ interacting via a non-local Hamiltonian $H_{AB}$ of the form:
\begin{equation}
    {H}_{AB}= J\left( \mu_{x}  \sigma^A_{x}\otimes\sigma^B_{x} + \mu_{y}  \sigma^A_{y}\otimes\sigma^B_{y} +\mu_{z}  \sigma^A_{z}\otimes\sigma^B_{z}\right),\label{equ:hamiltonian}
\end{equation}
where $J$ is a constant with the unit of energy, and $\mu_{x}$, $\mu_{y}$, $\mu_{z}$ are dimensionless with ordering $\mu_{x} \ge \mu_{y} \ge \mu_{z} \ge 0$. For unitary dynamics, Eq.~\eqref{Master_equation_for_density_operator} reduces to the Liouville-von Neumann equation
\begin{equation}
     \dot{\Psi_t}= -\frac{\iota}{\hbar}[H,\Psi_{t}],
\end{equation}
where $H$ is the Hamiltonian of the system. We take $\Psi_0 = \ketbra{\Psi_{0}}{\Psi_{0}}$ as initial state of the bipartite system $AB$ with $\ket{\Psi_{0}} = \sqrt{p}\ket{00}+\sqrt{1-p}\ket{11}$ for $p\in[0,1]$. The state $\rho_t$ of the system at the point of time $t$ is given by 
\begin{align}
   \Psi_t & = \frac{1}{2} \left(1+(2 p-1) \cos \left(2 \delta t\right)\right)\ketbra{00}{00} \nonumber\\
   & \hspace{0.35cm}+ (\sqrt{p(1-p) }+\frac{\iota}{2}  (2 p-1) \sin \left(2 \delta t\right)) \ketbra{00}{11} \nonumber\\
   & \hspace{0.35cm}+ (\sqrt{p(1-p)} + \frac{\iota}{2}  (1-2 p) \sin \left(2 \delta t\right)) \ketbra{11}{00} \nonumber\\
   & \hspace{0.35cm}+ \frac{1}{2} \left(1+(1-2 p) \cos \left(2 \delta t\right)\right) \ketbra{11}{11}, \label{equ:final_state_in_non-local_dynamics}
\end{align}
where $\delta:=\mu_{x} - \mu_{y}$. Here, $t = \frac{J \Tilde{t}}{\hbar}$ is the dimensionless parameter being used to measure time, with $\Tilde{t}$ being the actual time. The CSS~(with respect to REE) at $t=0$ is given by $\sigma^{*}_{0} = p \ketbra{00}{00}+(1-p)\ketbra{11}{11}$~\cite{Vedral1997}. Let $\mathcal{E}^{*}_{AB}$ denote the closest separable operation of the unitary $U_{AB}= {\rm e}^{-iH_{AB}t}$. The action of $\mathcal{E}^{*}_{AB}$ on $\sigma^{*}_{0}$ is given by
\begin{equation}
     \mathcal{E}^{*}_{AB} (\sigma^{*}_{0}) = \frac{1}{4}\left((x+2)\ketbra{00}{00}+(2-x)\ketbra{11}{11}\right),
\end{equation}   
where $x :=(2 p-1) \cos (2 \delta  t)+(2 p-1) \cos (2 \theta  t)$ and $\theta:=\mu_{x} + \mu_{y}$. 
To calculate bound \ref{equ:speeed_limit_entanglemen_unitary}, we have to calculate the following quantities:
\begin{align}
    \cal{F}_{E}(0) &= - p \log(p) - (1-p) \log(1-p),\nonumber\\
    \cal{F}_{E}(t) &= -\frac{1}{2} ((2 p-1) \cos (2 \delta  t)+1) \log \left(\frac{1}{4} (x+2)\right)\nonumber\\
    &\hspace{0.35cm}-\frac{1}{2} ((1-2 p) \cos (2 \delta  t)+1) \log \left(\frac{1}{4} (2-x)\right),\nonumber\\
     U_{\ln{\sigma_{t}}}& = -\frac{1}{2} \left((1-2 p)^2 \cos (4 \delta  t)+4 (p-1) p-1\right) y,\nonumber\\
     U_{H}& = \delta ^2 (1-2 p)^2,
\end{align}
where $ y := \tanh ^{-1}\left(\frac{x}{2}\right)^2 $  and 
\begin{equation}
   \left| \tr(\Psi_t G_{\sigma_t}(\Dot{\sigma}_t))\right| =  2 \left| \frac{g(p,t,\theta,\delta)}{(x-2)(x+2)}\right|,
\end{equation}
where $ g(p,t,\theta,\delta)$  is function of $p$, $t$, $\theta$, and  $\delta$, given as 
\begin{align}
   g(p,t,\theta,\delta)& := (1-2 p)^2 (\cos (2 t \delta )-\cos (2 t \theta )) \nonumber\\
   &\hspace{0.35cm}(\delta  \sin (2 t \delta )+\theta  \sin (2 t \theta )).
\end{align}
\begin{figure}
    \begin{center}
    \centering    
    \includegraphics[width=6.8cm]{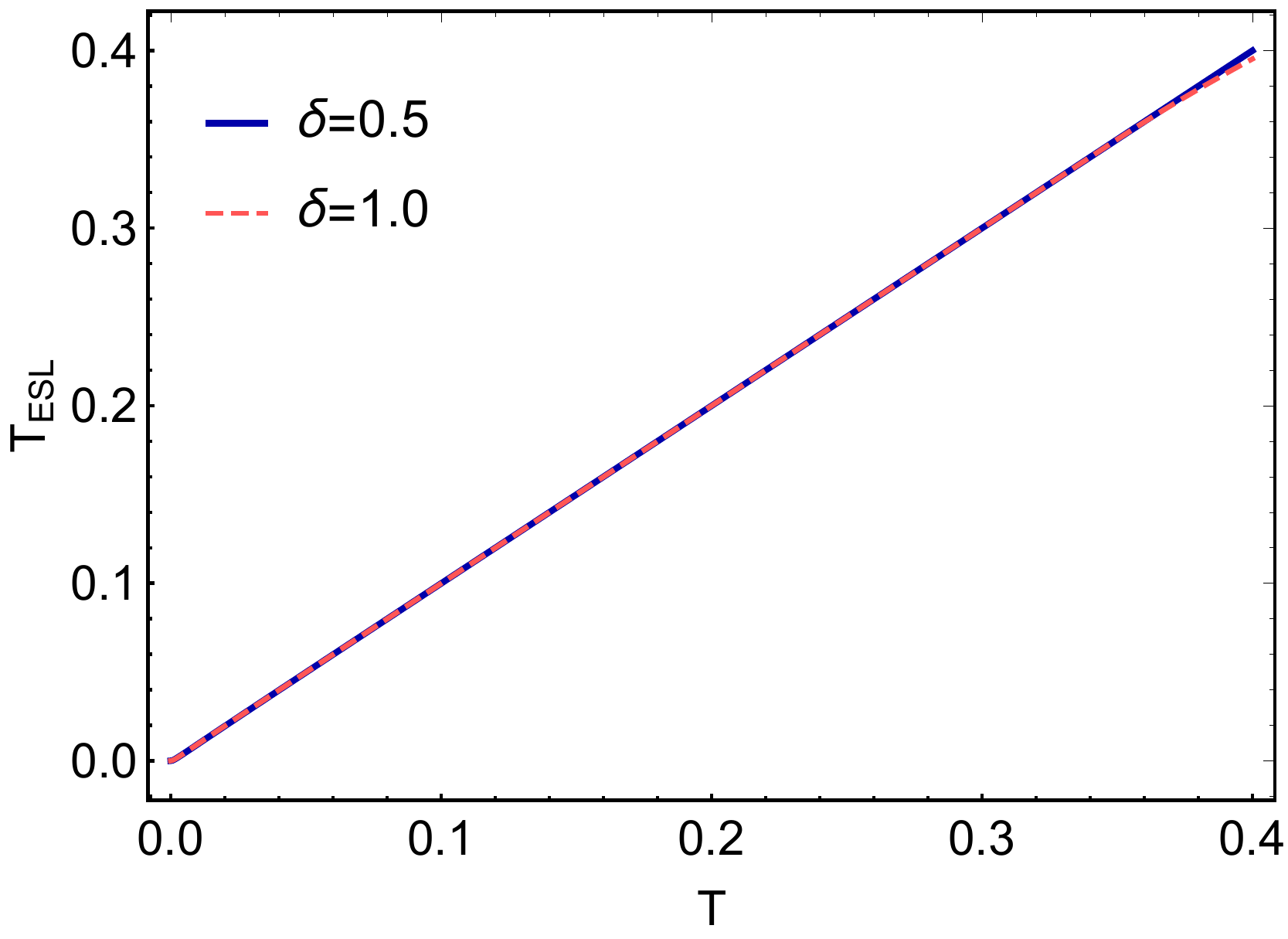}
    \caption{Here we depict $T_{\rm ESL}$ vs $T$, as given by~\ref{equ:speeed_limit_entanglemen_unitary}, for the unitary dynamics governed by the Hamiltonian given in Eq.~\eqref{equ:hamiltonian} with the pure input $\Psi_{0} = \sqrt{p}\ket{00}+\sqrt{1-p}\ket{11}$. We have taken  $p=0 $ and Hamiltonian parameters $\theta = 3.5$ and $\delta = 0.5$ (blue line), $\delta = 1.0$ (red line).}
  \label{fig:Pure_Dephasing}    
    \end{center}
\end{figure}
 In Fig.~\ref{fig:Pure_Dephasing}, we plot $T_{\rm ESL}$ vs $T$, as given by~(\ref{equ:speeed_limit_entanglemen_unitary}), for unitary dynamics generated by the two-qubit non-local Hamiltonian with Hamiltonian parameters $ \theta = 3.5$, and $\delta = 0.5$ (blue line), $\delta = 1.0$ (red line), and state parameter $p = 0$. For these parameters we have found that $T\approx T_{ESL}$ which suggest that bound~(\ref{equ:speeed_limit_entanglemen_unitary}) is tight
and attainable for considered unitary dynamics and initial states.

 Next, we demonstrate the tightness and attainability of bound~(\ref{equ:speeed_limit_entanglemen_unitary}) for mixed initial states and the bipartite systems are interaction via the non-local Hamiltonian given in Eq.~\eqref{equ:hamiltonian}.  We take $\rho_0 = p\ketbra{\Psi^{+}}{\Psi^{+}}+(1-p)\ketbra{00}{00}$ as initial state of the bipartite system $AB$ where $p\in[0,1]$ and $\ket{\Psi^{+}} = \frac{\ket{00}+\ket{11}}{\sqrt{2}}$ is a maximally entangled state. The state $\rho_t$ of the system at the point of time $t$ is given by 
\begin{align}
   \rho_t & =  \frac{1}{2}-\frac{1}{2} (p-1) \cos (2 t \delta )\ketbra{00}{00} \nonumber\\
   & \hspace{0.35cm}+\left(\frac{1}{2} (p-i (p-1) \sin (2 t \delta ))\ketbra{00}{11} +h.c. \right)\nonumber\\
   & \hspace{0.35cm}+ \frac{1}{2} ((p-1) \cos (2 t \delta )+1) \ketbra{11}{11}, \label{equ:final_state_mixed}
\end{align}
where $\delta=\mu_{x} - \mu_{y}$. The CSS~(with respect to REE) at $t=0$ is given by $\sigma^{*}_{0} = \left(1- \frac{p}{2} \right)\ketbra{00}{00}+\frac{p}{2}\ketbra{11}{11}$~\cite{Vedral1997}. Let $\mathcal{E}^{*}_{AB}$ denote the closest separable operation of the unitary $U_{AB}= {\rm e}^{-iH_{AB}t}$. The action of $\mathcal{E}^{*}_{AB}$ on $\sigma^{*}_{0}$ is given by
\begin{align}
     \mathcal{E}^{*}_{AB} (\sigma^{*}_{0}) &= \frac{1}{4} (2-(p-1) (\cos (2 t \delta )+\cos (2 t \theta )))\ketbra{00}{00}\nonumber\\
     &\hspace{0.35cm}+ \frac{1}{4} ((p-1) (\cos (2 t \delta )+\cos (2 t \theta ))+2)\ketbra{11}{11},
\end{align}   
where $\theta=\mu_{x} + \mu_{y}$. 

To calculate bound~(\ref{equ:speeed_limit_entanglemen_unitary}), we have to estimated $\mathcal{F}_{E}(t)$, $U_{\ln \sigma_t}, U_{H}$ and $\left| \tr(\rho_t G_{\sigma_t}(\Dot{\sigma}_t))\right|$ (see  Appendix~\ref{calculations_for_mixed_state}). In Fig.~\ref{mixedstate}, we have plotted the bound for pure and mixed initial states; for the pure product initial state, we observe that the bound saturates as $T \approx T_{ESL}$; however, for mixed states (mixture of product and maximally entangled state), it's slightly loose, particularly for the early time. This analysis suggests that the bounds are attainable for Higenberg interaction and considered initial states.

\begin{figure}
\centering    
     \includegraphics[width=7cm]{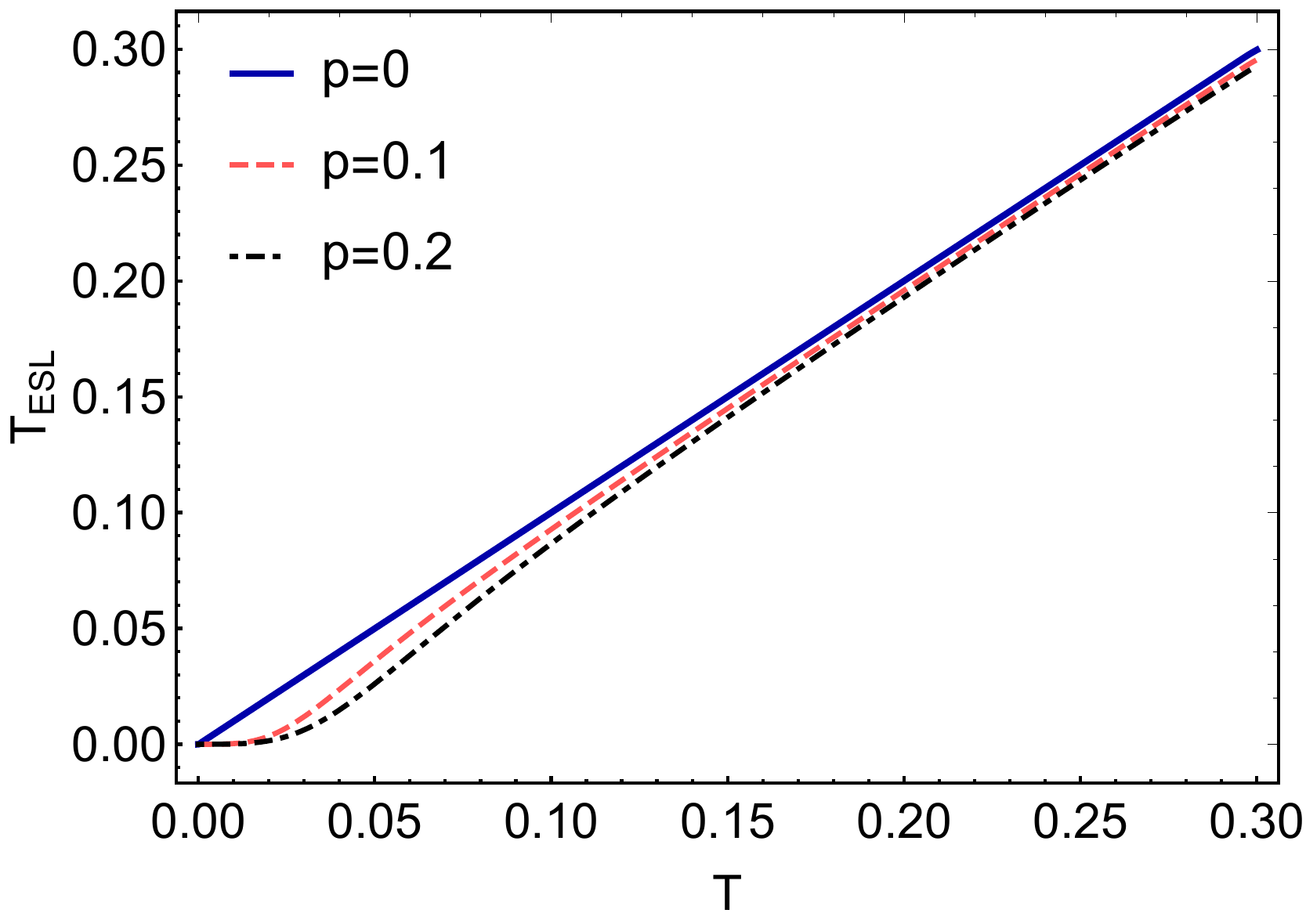}
    \caption{Here we depict $T_{\rm ESL}$ vs $T$, as given by~\ref{equ:speeed_limit_entanglemen_unitary}, for the unitary dynamics governed by the Hamiltonian given in Eq.~\eqref{equ:hamiltonian} with initial state $ p\ketbra{\Psi^{+}}{\Psi^{+}}+(1-p)\ketbra{00}{00}.$ For $p=0,0.1$ and $0.2$ we have taken $\theta = 3.5, \delta = 0$ (blue line), $\theta = 3.5, \delta = 0.1$ (red line) and $\theta = 5.5, \delta = 0.1$ (black line), respectively.}\label{mixedstate}
\end{figure}

{\it Open system dynamics}.--  For the open system dynamics, under the Markovian approximation, we have the following form of the master equation~\cite{Lindblad1976, Gorini1975}:

\begin{equation}
    \frac{\mathrm{d}\rho_{t}}{\mathrm{d}t} = -\iota[H,\rho_{t}]+\sum_{\alpha}\left(2 L_{\alpha}\rho_{t} L^{\dag}_{\alpha}-\{L^{\dag}_{\alpha}L_{\alpha},\rho_{t}\}\right) ,
\end{equation}
where $L_{\alpha}$'s  are called the Lindbladian or quantum jump operators, $H$ is the driving Hamiltonian of the system and $\{A_1,A_2\}:= A_1A_2 + A_2A_1$ denotes the anti-commutator bracket. The above equation is called the Gorini-Kossakowski-Lindblad-Sudarshan master equation. 

 We consider a bipartite system $AB$, where the subsystems $A$ and $B$ are qubits, each coupled with environments $E_{A}$ and $E_{B}$, respectively, where $E_A$ and $E_B$ are not interacting with each other. We assume that the system $AB$ is initialised in a pure state  $\rho_0$ of the form
\begin{align}
     \rho_0 &=p \ketbra{00}{00} + \sqrt{\left(1-p\right) p}(\ketbra{00}{11}+\ketbra{11}{00}) \nonumber \\
      & \hspace{0.35cm}+ \left(1-p\right) \ketbra{11}{11},
\end{align}
 where $p\in[0,1]$. The CSS~( with respect to REE) of the above state is given by $\sigma^{*}_{0} = p \ketbra{00}{00}+(1-p)\ketbra{11}{11}$~\cite{Vedral1997}.
 
 We consider a pure dephasing channel as an example of a quantum correlation degradation process. The quantum jump operators for a pure dephasing process are given as $L_{1} = \sqrt{\frac{\gamma^A}{2}} \sigma^A_{z}\otimes \mathcal{I}_{B}$ and $L_{2} = \sqrt{\frac{\gamma^B}{2}}\mathcal{I}_{A} \otimes\sigma^B_{z} $, where $\sigma^A_{z}$ and $\sigma^B_{z}$ are Pauli operators acting on systems $A$ and $B$, respectively, and $\gamma^A,\gamma^B$ both are real parameters denoting the strength of the dephasing process. The master equation governs time-evolution of the state $\rho_{t}$, and is given by
 \begin{align}\label{rho:dep}
  \frac{\rm d}{{\rm d}t}\rho_{t} &= \gamma^A\left(\sigma^A_{z}\otimes \mathcal{I}_{B}\left(\rho_{t}\right)\sigma^A_{z}\otimes \mathcal{I}_{B} -\rho^{AB}_{t}\right)\nonumber\\
  & \hspace{0.35cm}+\gamma^B\left( \mathcal{I}_{A}\otimes\sigma^B_{z}\left(\rho^{AB}_{t}\right)\mathcal{I}_{A}\otimes\sigma^B_{z} -\rho_{t}\right).
   \end{align}
The state of the quantum system $AB$ at time $t$ is given by 
  \begin{align}
      \rho_{t} &= p \ketbra{00}{00} +\sqrt{p\left(1-p\right)} {\rm e}^{-4 \gamma  t}\left( \ketbra{00}{11}+\ketbra{11}{00}\right) \nonumber \\
      & \hspace{0.35cm}+ \left(1-p\right) \ketbra{11}{11}
  \end{align}
  where we have assumed that the dephasing rates of both environments are equal to $\gamma$. The CSS at $t=0$ is given by $\sigma^{*}_{0} = p \ketbra{00}{00}+(1-p)\ketbra{11}{11}$. The pure dephasing process is a separable operation, so that the operation itself is its closest separable operation. Now, the CSS at time $t$ is given by  
  \begin{equation}
      \sigma^{*}_{t} = p \ketbra{00}{00}  \left(1-p\right) \ketbra{11}{11}.
  \end{equation}

 To calculate the bound~\ref{equ:speeed_limit_entanglemen}, we have to calculate the following quantities
\begin{align}
     \cal{F}_{E}(0) &= - p \log(p) - (1-p) \log(1-p),\nonumber\\
    \cal{F}_{E}(t) &= h(p,\gamma,t).\\
    \left| \tr(\rho_t G_{\sigma_t}(\Dot{\sigma}_t))\right| & = 0,\\
    \norm{\mathcal{L}_{t}({\rho_t})}_{\rm 2}& = 4 \sqrt{2} \sqrt{\gamma ^2 (1-p) p e^{-8 \gamma  t}},
\end{align}
where $h(p,t,\gamma)$ is function of $p$, $t$ and  $\gamma$. The analytic expression of $h(p,t,\gamma)$ is given in Appendix~\ref{appendix:pure_dephasing_process}.
We have calculated $ \norm{\ln \rho_t-\ln \sigma_t}_{\rm 2}$ for  $p= 0.5, 0.4 $ and $0.3$. All the expressions are given in the Appendix~\ref{appendix:pure_dephasing_process}.

\begin{figure}
  \includegraphics[width=7cm]{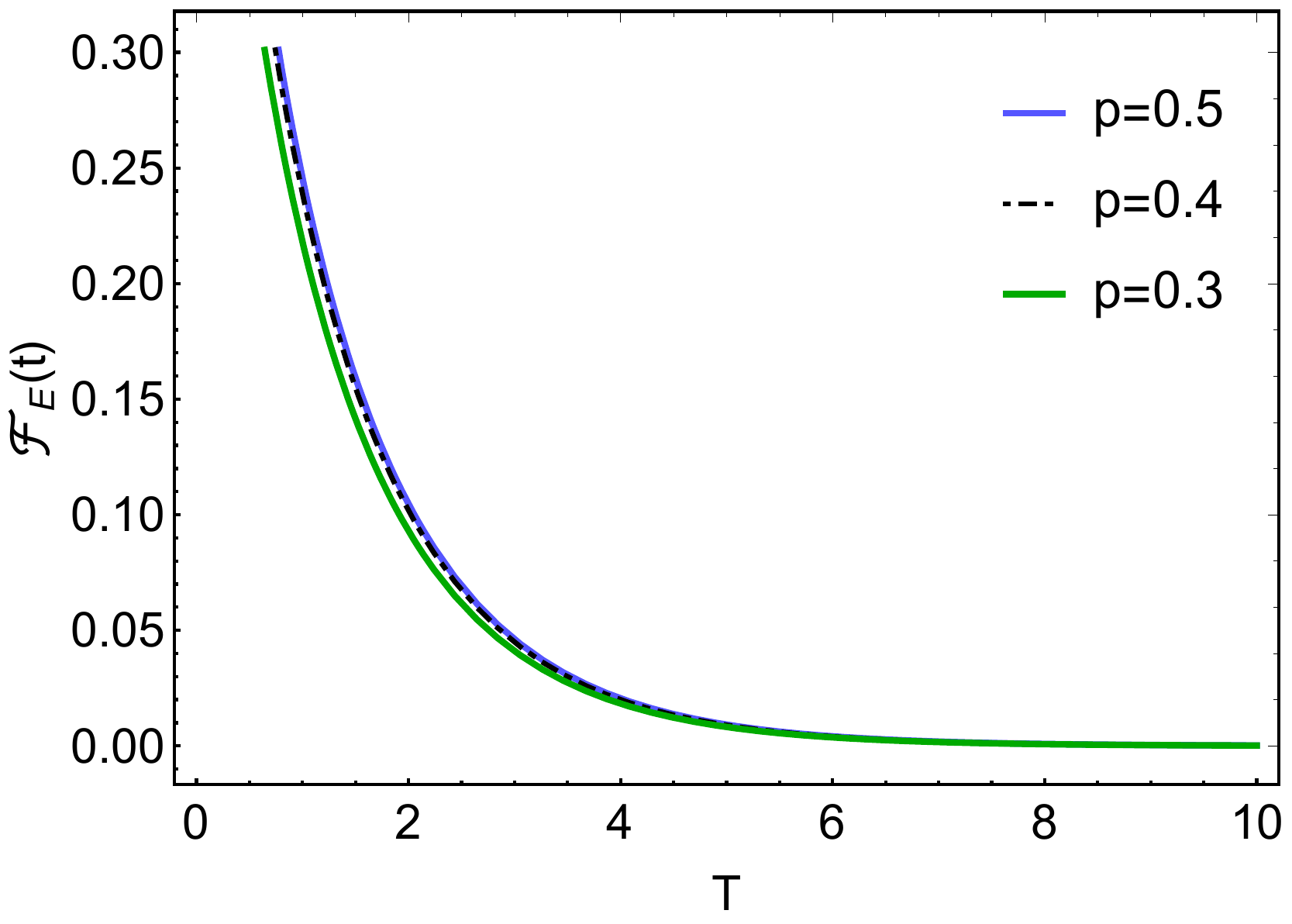}
     \includegraphics[width=7cm]{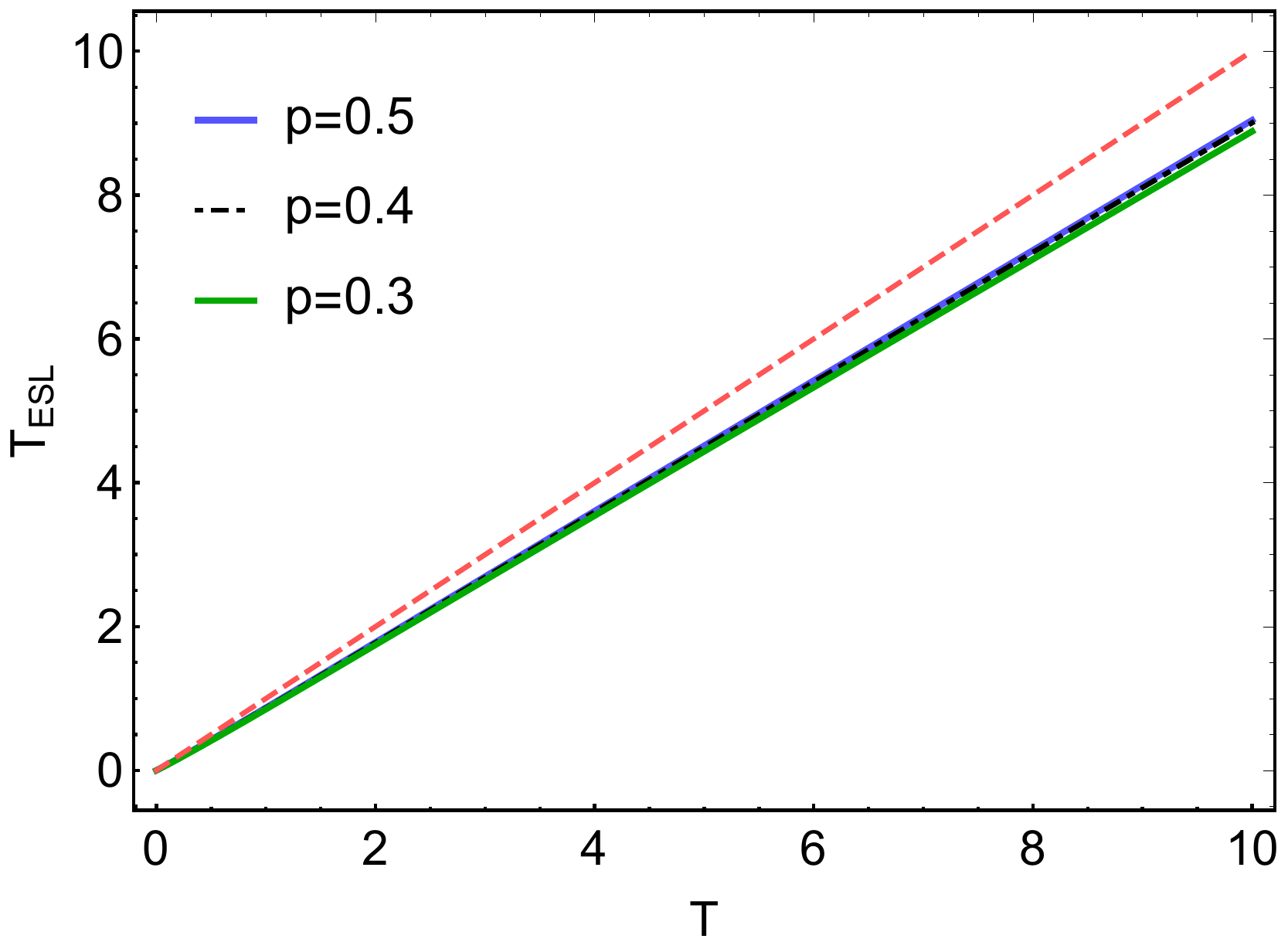}
    \caption{Here we depict $T_{\rm ESL}$ vs $T$, as given by~\ref{equ:speeed_limit_entanglemen}, for the pure dephasing process. We have taken the pure input state $\Psi_{0}=\sqrt{p} \ket{00}+\sqrt{1-p}\ket{11}$, and we have taken state parameter $p=0.5,0.4$ and $0.3$ and decay rate $\gamma = 0.1$.}\label{open_plot}
  
\end{figure}

In Fig.~\ref{open_plot}, we plot the relative entropy of entanglement and Bound~eq. for local dephasing dynamics for various initial pure entangled states. In the first figure, we observe that entanglement degrades as time progresses and reaches almost negligible for a long time. In the second figure, we demonstrated that considering the entangled state and dephasing dynamic, the bound is initially very tight, and later time, it becomes slightly loose as the state becomes mixed. Thus, we see that for considered dephasing dynamics, the bound can be almost attainable.

\section{conclusion}\label{conclusion}
We have derived distinct speed limits on entanglement for both unitary and arbitrary dynamics using different measures of entanglement. Our approach relied on the assumption that the dynamics of the closest separable state can be described, or at least ``faithfully'' mimicked by the closest separable dynamical map of a given dynamical map. 
 We have found that the upper bound on the entanglement rate of bipartite quantum systems depends on two key factors: the system's evolution speed and a time-dependent term (due to the evolution of the closest separable state that applies to both unitary and arbitrary completely positive and trace preserving (CPTP) dynamics. The speed limits on entanglement are fundamentally different from traditional speed limits of state evolution, as it is based on the rate of change of entanglement between subsystems rather than the rate of change in the distinguishability of the initial and evolved states of a system. It is important to identify, understand, and possibly exploit this difference, given that entanglement plays a critical role in quantum information processing, quantum communication, and quantum sensing, and its efficient manipulation and generation is key to the development of quantum technologies. By setting a fundamental limit on the rate at which entanglement can be generated or manipulated in bipartite systems, these speed limits provide a tool for evaluating the efficiency of different physical processes for generating and manipulating entanglement. Moreover, the speed limits on entanglement determine the minimum time required for certain changes in the amount of entanglement of a bipartite system using any physical process. The speed limits on entanglement are closely related to entangling rates, which have been extensively studied for closed~\cite{Vidal2001, marien2016, Vershynina_2019, Gong2022, Hamilton2023} and open bipartite systems~\cite{Anna2015}. We found that the speed limits on entanglement for a two-qubit system governed by a non-local Hamiltonian are both tight and attainable. Our method can be used to find speed limits for other quantum resources and correlations as well (See Appendix~\ref{speed_limit_on_quantum_resources}).
 
\section*{Acknowledgments}
 VP and SB acknowledge the support of the INFOSYS scholarship. US acknowledges partial support from the Interdisciplinary Cyber-Physical Systems (ICPS) program of the Department of Science and Technology (DST), Government of India, Grant No. DST/ICPS/QuST/Theme3/2019/120

\appendix

\section{Detailed Preliminaries} \label{prelims}

The Schatten-p norm of an operator $ {\cal{O}} \in \cal{L}(\cal{H})$ is defined as
\begin{equation}
       \norm{{\cal{O}}}_{p}= \left(\tr \left|{\cal{O}}\right|^{p}\right)^{1/p},
\end{equation}
where $\left|{\cal{O}}\right|= \sqrt{{\cal{O}}^{\dagger}{\cal{O}}}$, $p\geq1,\ p\in\mathbb{R}$.  The operator norm, the Hilbert-Schmidt norm, and the trace norm correspond to $p=\infty,2,1$ respectively and satisfy the inequality $\norm{A}_{\rm \infty} \leq \norm{A}_{\rm 2}   \leq \norm{A}_{\rm 1} $.

{\it Quantum dynamics}.-- For any dynamical quantum system with associated Hilbert space $\cal{H}$, the state of the system, generically,  changes with time. The time evolution of the state of the system is given by a dynamical map which maps an initial state to a final state and it should satisfy the following properties:\\
$(i)$ {\it Linearity}.-- A  map $\cal{M}:\cal{L}({\cal{H}}_{A})\rightarrow\cal{L}({\cal{H}}_{B})$ is called linear if ${\cal{M}}(a X_{A}+ b Y_{A})
= a \Phi(X_{A})+b \Phi(Y_{A})$, where $X_{A},Y_{A} \in {\cal{L}(\cal{H}}_{A})$ and $a,b \in \mathbb{C}$. $(ii)$ {\it Complete Positivity }.--  A linear map $\cal{M}:\cal{L}({\cal{H}}_{A})\rightarrow\cal{L}({\cal{H}}_{B})$ is called
positive if
$\cal{M}\left(\eta_{A}\right)\in\cal{L}({\cal{H}}_{B})_{+}~\forall~\eta_{A}~\in \cal{L}({\cal{H}}_{A})_{+}$ and completely positive if  $\rm id_{R}\otimes\cal{M}$ (where $\rm id_{R}$ denotes the identity super-operator acting on an arbitrary reference system $R$) is also a positive map. $(iii)$ {\it Preservation of Trace}.-- A map $\cal{M}:\cal{L}({\cal{H}}_{A})_{+}\rightarrow\cal{L}({\cal{H}}_{B})_{+}$ is called trace preserving  if $\tr\left(\cal{M}(\eta_{A})\right)=\tr(\eta_{A})~\forall~\eta_{A}\in{\cal{L}}({\cal{H}}_{A})$.

A mapping which fulfils the above three conditions is called a quantum channel (dynamical map) and it describes a physically meaningful evolution of density operators. Sometimes, it could be easier to use differential equations to describe the evolution instead of quantum
channels:
\begin{equation}
      \dot{\rho_t}:=\diff{\rho_t}{t}=\mathcal{L}_t(\rho_t) \label{Master_equation_for_density_operator},
 \end{equation}
where $\mathcal{L}_t$ is the Liouvillian super-operator~\cite{Rivas2012} which in general can be time independent or time-dependent.

{\it von Neumann entropy}.--The von Neumann entropy of a quantum state $\rho_A \in \cal{S}(\cal{H}_A)$ is a non-negative quantity given by:
 \begin{equation}
     S(A)_{\rho}:=S(\rho_A)= -\tr({\rho_{A}\log{\rho_{A}}}),
 \end{equation}
 where the logarithm is taken to base 2.
  $S(\rho_{A})$ is manifestly invariant under a unitary transformation $\rho_{A}\rightarrow U\rho_{A} U^{\dag}$, i.e., $S(\rho_A)=S(U\rho_{A} U^{\dag})$.
  
{\it Relative entropy}.-- The quantum relative entropy between density operators $\rho \in \cal{S}(\cal{H})$ and a positive semi-definite operator $\sigma\in\cal{L}(\cal{H})$ is defined as follows \cite{Umegaki1962}
 \begin{equation}
    D(\rho\Vert\sigma) := 
       \begin{cases}
                \tr(\rho(\log{\rho}-\log{\sigma})) &  \text{if}\  \operatorname{supp}(\rho)\subseteq \operatorname{supp}(\sigma), \\ 
                +\infty & \text{otherwise},
       \end{cases}\label{equ:relative entropy}
 \end{equation}
where $\operatorname{supp}(\rho)$ and $\operatorname{supp}(\sigma)$ are the supports of $\rho$ and $\sigma$, respectively. The quantum relative entropy has a number of useful properties. $(i)$   $D(\rho\Vert\sigma) $ is invariant under unitary dynamics, i.e., $D(\rho\Vert\sigma)=D(U\rho U^{\dag}\Vert U \sigma U^{\dag})$, and non-increasing under the action of linear CPTP maps, i.e., $ D(\rho\Vert\sigma) \geq  D({\cal{N}}(\rho)\Vert{\cal{N}}(\sigma)) $ for any density operators $\rho$ and $\sigma$ and for any linear CPTP map $\cal{N}$. $(ii)$ Relative entropy between two positive semi-definite operators $\rho$ and $\sigma$ is positive if $\rho\in\cal{S}(\cal{H})$, $\sigma\in{\cal{L}}'(\cal{H})_{+}$ and $\tr(\sigma)\leq 1$, with equality if and only if $\sigma=\rho$. So, quantum relative entropy can also be thought of as an appropriate measure of distinguishability or discrimination between $\rho$ and $\sigma$, which motivates one to think $D(\rho\Vert\sigma)$ as a kind of ``metric" (a measure of distance) between density operators. Note that $D(\rho\Vert\sigma)$ is not a true metric because it is neither symmetric between $\rho$ and $\sigma$ nor it satisfies triangular inequality, but it is an entropic measure of distance between two density matrices.

{\it Uncertainty relation for mixed states}.-- For any two observables $\{{\cal{O}}_{1}$, ${\cal{O}}_{2}\} \in {\cal{L}}(\cal{H})$, and $\rho\in{\cal{S}}(\cal{H})$, the following inequality holds~\cite{Luo2005}
\begin{align}
    \frac{1}{4} \left|\tr\left([{\cal{O}}_1,{\cal{O}}_2]\rho\right)\right|^2\leq U_{{\cal{O}}_1} U_{{\cal{O}}_2}, \label{equ:uncertainty_relation_for_mixed_states}
\end{align}
where $U_{\cal{O}} = \sqrt{V^{2}_{\cal{O}}-(V_{\cal{O}}-I_{\cal{O}})^2}$ for an operator ${\cal{O}}\in{\cal{L}}(\cal{H})$, $V_{\cal{O}}=\tr(\rho{\cal{O}}^2)-(\tr(\rho{\cal{O}}))^2$ is variance of the operator $\cal{O}$ in the state $\rho$, and $I_{\cal{O}}$ is Wigner-Yanase skew information, defined as :
\begin{equation}
    I_{\cal{O}} = -\frac{1}{2}\tr\left([\sqrt{\rho},{\cal{O}}]^2\right).
    \end{equation}
One can easily check that the above uncertainty reduces to well known Robertson uncertainty relation~\cite{robertson_uncertain}, when $\rho$ is a pure state.

{\it CJ isomorphism}.--  
Let $\mathcal{H}_R$ be a Hilbert space isomorphic to another Hilbert space $\mathcal{H}_A$ and let their orthonormal bases be $\{ \ket{i}_R \}$ and $\{ \ket{i}_A \}$ respectively. Now the space $\mathcal{L}(\mathcal{L}(\mathcal{H}_{A}),\mathcal{L}({\mathcal{H}}_{B}))$ of linear maps  $\mathcal{E}:{\mathcal{L}(\mathcal{H}}_{A})\rightarrow\mathcal{L}({\mathcal{H}}_{B})$ and the tensor product space  $\mathcal{L}(\mathcal{H}_{R}) \otimes \mathcal{L}(\mathcal{H}_{B}) $ are isomorphic under the following identification:

\begin{align}
\mathcal{E} \leftrightarrow
    \Phi_{RB} & \equiv ({\rm id}_R \otimes \mathcal{E})\left(\left(\ketbra{\Psi}{\Psi})_{RA}\right)\right)\nonumber\\
    &= \sum_{i,j} \ketbra{i}{j}_R \otimes \mathcal{E} (\ketbra{i}{j}_A)
\end{align}
where $\ket{\Psi}= \sum_i \ket{i}_R\otimes \ket{i}_A$ is an unnormalized maximally entangled state of $RA$. The $\mathcal{E} \leftrightarrow  \Phi_{RB}$ correspondence is known as the Choi-Jamio{\l}kowski(CJ) isomorphism and the operator $ \Phi_{RB}$ is known as the {\it Choi operator} of the linear map $\mathcal{E}$. Choi's theorem for a completely positive map guarantees that $\Phi_{RB}$ is positive semi-definite if and only if $\mathcal{E}$ is completely positive. Similarly a CP map $\mathcal{M}_{AB}$ on a bipartite system $AB$ can have a Choi operator representation given by
\begin{equation}
    \Phi_{AA'BB'}=({\rm id}_{A'B'}\otimes \mathcal{M}_{AB})(\Psi_{AA'}\otimes \Psi_{BB'}),
\end{equation}
where $\mathcal{H}_A'$ and $\mathcal{H}_B'$ are isomorphic to $\mathcal{H}_A$ and $\mathcal{H}_B$ respectively. $\Psi_{AA'}$ and $\Psi_{BB'}$ are the maximally entangled states on $\mathcal{H}_{AA'}$ and $\mathcal{H}_{BB'}$ respectively. Conversely, given a positive semi-definite operator on $AA'BB'$, the corresponding CP map is given by:
\begin{equation}
    \mathcal{M}(\rho_{AB})=d^2 \tr_{A'B'}(\Phi_{AA'BB'}\rho^T_{A'B'}\otimes \mathcal{I}_{AB})
\end{equation}
where $T$ denotes transposition with respect to the chosen basis of $\mathcal{H}_{A'}\otimes \mathcal{H}_{B'}$ and we have taken the dimensions of all the systems to be $d$. 
Using the above relations, it can be shown that \cite{Cirac_2001}:
\begin{itemize}
    \item $\mathcal{M}$ is separable iff $\Phi_{AA'BB'}$ is separable with respect to the systems $\mathcal{H}_{AA'}$ and $\mathcal{H}_{BB'}$.
    \item $\mathcal{M}$ can create entanglement iff $\Phi_{AA'BB'}$ is non separable with respect to the subsystems $\mathcal{H}_{AA'}$ and $\mathcal{H}_{BB'}$.
    \item If $\mathcal{M}$ corresponds to a unitary channel, then its Choi operator has rank one, i.e., it can be written as a projector: $\Phi_{AA'BB'}= \ketbra{\phi}{\phi}$, where $\ket{\phi}$ is a state in $\mathcal{H}_{AA'}\otimes \mathcal{H}_{BB'}$.
\end{itemize}

\section{ Derivative of \texorpdfstring{$f(t) = \tr(\rho_t\log \sigma_t) $}{}}\label{derivation_of_D(f(t))}
We want to calculate derivative of the function $f(t) (=\tr(\rho_t\log \sigma_t))$ with respect time $t$. As we have assumed that the derivative of $D\left(\rho_{t}\Vert\sigma_{t}\right)$ is well defined in $t\in I$, this also implies that  $f(t) $ is a smooth function of $t$~(i.e., $f(t)$ is continuous and differentiable) in the interval $I$. Then, $\frac{\rm d}{{\rm d}t} f(t)$ can be written as 
\begin{equation}
    \frac{\rm d}{{\rm d}t}f(t) = \lim_{\epsilon\rightarrow 0}\frac{f(t+\epsilon)-f(t)}{\epsilon}.\label{derivative}
\end{equation}
where $\epsilon$ is a small positive real number.
The term $f(t+\epsilon)$ can be calculated by Taylor expansion ( after ignoring the higher order terms in $\epsilon$)
\begin{align}
    f(t+\epsilon) &=\tr\left((\rho_t+\epsilon\Dot{\rho}_t) \log(\sigma_t+\epsilon \Dot{\sigma}_t)\right) \nonumber\\
    &= \scalemath{0.95}{\tr\left(\rho_t\log(\sigma_t+\epsilon \Dot{\sigma}_t))+\epsilon\tr(\Dot{\rho}_t\log(\sigma_t+\epsilon \Dot{\sigma}_t)\right)}. \label{equ:taylor_expansion_of_f}
\end{align}

We now calculate $\log(\sigma_t+\epsilon \Dot{\sigma}_t)$ separately. Consider a small real number q in the $\delta$ neighbourhood of $0$, i.e., $q\in(-\delta,\delta)$, and two hermitian operators $A\in {\cal{L}}({\cal{H}})_{+}$, $B\in{\cal{L}}({\cal{H}})$ such that operator $A$ is full rank, then the Rellich's theorem~\cite{kato2013perturbation} says that $\log(A+qB)$ will be analytic in $q$. We then have the Taylor expansion
\begin{equation}
    \log(A+qB)=\log A+ q G_{A}(B)+ O(q^2),\label{equ:Taylor_exp_of_log}
\end{equation}
where $G_A(B)$ is the following self-adjoint linear operator:
Working in the eigenbasis of $A=SMS^{-1}$, where $M = {\rm diag}(a_1,a_2,..a_{{\rm dim(\cal{H})}})$ is a diagonal matrix with eigenvalues of $A$ as its elements, define the matrices:
\begin{eqnarray}
    T_{kl}(A)= \frac{\log a_k - \log a_l}{a_k - a_l} \label{equ:T}\\
    S_{ij}(A)= \frac{a_i- a_j}{\log a_i - \log a_j}.\label{equ:S}
\end{eqnarray}
Here we assume that $\frac{\log a -\log a}{a-a}=\frac{1}{a}$. Then, $G_{\alpha}(\beta)=\beta \circ T(\alpha)$, where $A \circ B$ is the component-wise product of A and B, also known as Hadamard product. Note that in the case of singular $A$, Eq.~\eqref{equ:Taylor_exp_of_log} is not well defined but the following still holds~\cite{Fried2011}
\begin{equation}
   \scalemath{0.90}{\tr(C\log(A+q B))= \tr ( C \log A) + q \tr(C G_A(B))+ O(q^2)}, \label{equ:logexpansion}
\end{equation}
assuming that $\operatorname{supp}(C)\subseteq \operatorname{supp}(A)$ for some $C \in {\cal{L}}(\cal{H})$ and operator $ A+qB$ is full rank. In this scenario, matrix elements of  $T$ and $S$ are defined the same as above~(Eqs.~\eqref{equ:T} and~\eqref{equ:S}) on the support of $A$ and zero outside the support. If $A+qB$ is not full rank, but $\operatorname{supp}(C)\subseteq \operatorname{supp}(A+qB)$, then Eq.~\eqref{equ:logexpansion} can be modified as follows
\begin{align}
  \tr(C\Pi_{A+qB}\log(A+q B)) = & \tr ( C \Pi_{A+qB}\log A)\nonumber\\
  &\scalemath{0.92}{+ q \tr(C\Pi_{A+qB} G_A(B))+ O(q^2)} \label{equ:logexpansion1},
\end{align}
where $\Pi_{A+qB}$ is projection onto the support of operator $A+qB$. Since we are only interested in the case where $A$ is either a density operator or derivative of a density operator, $C$ is a density operator, and $B$ is a derivative of $C$. The above equation implies for $A =\sigma_t$, $C=\rho_t$, $B=\dot{\sigma}_t$ and $q=\epsilon$
\begin{align}
    \tr(\rho_{t}\Pi_{\sigma_{t}}\log(\sigma_t+\epsilon \Dot{\sigma}_t))=&\scalemath{0.93}{ \tr ( \rho_{t} \Pi_{\sigma_{t}}\log \sigma_t)+ \epsilon \tr(\rho_{t} \Pi_{\sigma_{t}}G_{{\sigma}_t}(\dot{\sigma}_t))}\nonumber\\
    &+ O(\epsilon^2), \label{logexp}
\end{align}
where we have used the fact that for finite-dimensional Hilbert space, the support of $\Dot{\sigma_{t}}$ is contained in $\sigma_{t}$, so instead of $\Pi_{\sigma_{t}+\epsilon \Dot{\sigma_{t}}}$ we have used $\Pi_{\sigma_t}$. We have further assumed that $\dot{\sigma}_t$ is well-defined ~(by this, we mean that each matrix element of $\sigma_t$ is differentiable with respect to $t$). The above equation gives the first term on the right-hand side of Eq.~\eqref{equ:taylor_expansion_of_f}. To calculate the second term, we take $A =\sigma_t$, $C=\Dot{\rho_t}$, $B=\dot{\sigma}_t$ and $q=\epsilon$ in Eq.~\eqref{equ:logexpansion1}, we then get
\begin{align}
    \epsilon \tr(\Dot{\rho_{t}}\Pi_{\sigma_{t}}\log(\sigma_t+\epsilon \Dot{\sigma}_t))=& \epsilon \tr ( \rho_{t} \Pi_{\sigma_{t}}\log \sigma_t)+ O(\epsilon^2)\label{logexp1}.
\end{align}
 From Eqs.~\eqref{derivative},~\eqref{equ:taylor_expansion_of_f},~ \eqref{logexp}, and ~\eqref{logexp1}, we obtain~( after taking the limit $\epsilon \rightarrow 0 $ and  ignoring the terms of $O(\epsilon^2)$):
\begin{equation}
    \frac{{\rm d}}{{\rm d}t} f(t)= \tr (\Dot{\rho_t}{\Pi}_{\sigma_t}\log \sigma_t) + \tr(\rho_t {\Pi}_{\sigma_t}G_{\sigma_t}(\Dot{\sigma}_t)). \label{derivative_of_f}
\end{equation}

\section{Saturation of bound~(\ref{ETD})}\label{Sat}
If the closest separable state does not change in time, the bound~(\ref{ETD}). reduces to the following bound
\begin{align}~\label{TD}
    T \ge T^{tr}_{\rm ESL} := \frac{|{\cal{D}}_{tr}(\rho_T,\sigma_0)-{\cal{D}}_{tr}(\rho_0,\sigma_0)|}{\Lambda_{tr}(T)},
\end{align}
where $\Lambda_{tr}(T):=\frac{1}{T}\int_{0}^{T}  
    \tr|\Dot{\rho_t}|{\rm d}t$.
    
Next, we show that the above bound saturates when the system along the geodesic, which is given as (see Refs.~\cite{Deffner2017, Connor2023})
\begin{equation}\label{GD}
    \rho(t)= p(t)\rho(0)+ (1-p(t))\rho(T),
\end{equation}
where $0\leq p(t) \leq 1$.
The inequality~(\ref{TD}) is the consequence of the following inequality:
\begin{equation}
  \frac{d}{dt}  \norm{\rho(t) -\sigma(0)}_{tr} \leq \norm{\dot{\rho}(t)}_{tr}.
\end{equation}
Using Eq.~\eqref{GD} in the above inequality, we obtain the following equality
\begin{equation}
     \norm{\dot{p}(t)(\rho(0) -\rho(T))}_{tr} = |\dot{p}(t)| \norm{\rho(0) -\rho(T))}_{tr},
\end{equation}
when $\dot{p}(t)\geq 0$ and $\rho(T)=\sigma(0)$ (which correspond to the scenario when the final state is separable), that leads to saturation of the bound~(\ref{ETD}) under the considered assumption.

\section{ Speed limits on general quantum resources}\label{speed_limit_on_quantum_resources}
A similar analysis can also be done for other quantum resources if we take relative entropy as a measure of distance, then the resource value of a given state $\rho$ is given as~\cite{Chitambar2019}:
\begin{align}
    R(\rho) =&  \min_{\sigma \in F } D(\rho \Vert \sigma)\label{equ:resource}\\
    & = D(\rho \Vert \sigma^*),
\end{align}
where $F$ is the set of free states~(i.e., states with zero resource value) and $\sigma^*\in F$ is a state which realizes the minimum of Eq.~\eqref{equ:resource}, called closest free state. Now, for that given dynamics of $\rho$, we have to find its closest dynamics, which maps a set of free states to free states. For the given initial state, if the closest free state is known, then the maximal rate of resource variation and speed limit bound on the resource can be calculated in a similar way~(as we have calculated for relative entropy of entanglement). In this case, bounds on resource similar to \eqref{speeed_limit_entanglement_production} and \eqref{speeed_limit_entanglement_depletation} will provide a lower bound on the minimum time system will take to create (starting from the free state) or fully deplete a certain amount of resource. We note that speed limits on resources for arbitrary dynamics have been studied earlier using relative entropy as resource measure~\cite{Campaioli_2022}. For any dynamical system with associated Hilbert space $\cal{H}$, the state of the system depends on time. So, in general, the closest free state will also be time-dependent. The speed limit bounds in \cite{Campaioli_2022} are obtained on the assumption that during the entire evolution period, the closest free state is not evolving. In our analysis, we assume that the evolution of the closest free state is governed by the closest free operation (map)~(closest to the actual operation (map) governing the evolution of the state).

\onecolumngrid     

\section{Expression of required quantities to estimate the bound for unitary dynamics of mixed states} \label{calculations_for_mixed_state}
The relative entropy function $\mathcal{F}_{E}(t)$ is given by 
\begin{align}
    \mathcal{F}_{E}(t) &= \frac{1}{2} \left(1-\sqrt{2 p^2-2 p+1}\right) \log \left(\frac{1}{2} \left(1-\sqrt{2 p^2-2 p+1}\right)\right)+\frac{1}{2} \left(\sqrt{2 p^2-2 p+1}+1\right) \log \left(\frac{1}{2} \left(\sqrt{2 p^2-2 p+1}+1\right)\right)\nonumber\\
    &-\frac{1}{2} ((p-1) \cos (2 \delta  t)+1) \log \left(\frac{1}{4} ((p-1) \cos (2 \delta  t)+(p-1) \cos (2 \theta  t)+2)\right)\nonumber\\
    &-\left(\frac{1}{2}-\frac{1}{2} (p-1) \cos (2 \delta  t)\right) \log \left(\frac{1}{4} (-(p-1) \cos (2 \delta  t)-(p-1) \cos (2 \theta  t)+2)\right).
    \end{align}
Now to calculate $U_{\ln \sigma_t}$ and $U_{H}$, we need to calculate  $V_{\ln \sigma_t}, I_{\ln \sigma_t}$ and $V_{H}, I_{H}$, which are give as
\begin{align}
    V_{\ln \sigma_t} &= \left((p-1)^2 \cos ^2(2 \delta  t)-1\right) \left(-\tanh ^{-1}\left(\frac{1}{2} (p-1) (\cos (2 \delta  t)+\cos (2 \theta  t))\right)^2\right),\\
    I_{\ln \sigma_t} &= \frac{\left(\sqrt{2} \sqrt{-(p-1) p}-1\right) \left((p-1)^2 \cos (4 \delta  t)+p (2-3 p)-1\right) \tanh ^{-1}\left(\frac{1}{2} (p-1) (\cos (2 \delta  t)+\cos (2 \theta  t))\right)^2}{4 (p-1) p+2},  \\  
        V_{H} &= \delta ^2 \left(-\left(p^2-1\right)\right),\\
        I_{H} &= -\frac{\delta ^2 (p-1)^2 \left(\sqrt{2} \sqrt{-(p-1) p}-1\right)}{2 (p-1) p+1}.   
     \end{align}
We also need $ \left| \tr(\rho_t G_{\sigma_t}(\Dot{\sigma}_t))\right|$ which is given by 
\begin{equation}
    \left| \tr(\rho_t G_{\sigma_t}(\Dot{\sigma}_t))\right| =   2 \abs{ \frac{(p-1)^2 (\cos (2 t \delta )-\cos (2 t \theta )) (\delta  \sin (2 t \delta )+\theta  \sin (2 t \theta ))}{(p-1)^2 (\cos (2 t \delta )+\cos (2 t \theta ))^2-4}}.
\end{equation}

\section{Expression of required quantities to estimate the bound for dephasing dynamics} \label{appendix:pure_dephasing_process}

The function $ h(p,t,\gamma)$ is given as 
\begin{align}
h(p,t,\gamma) &=-\frac{1}{2} \log \left(-\frac{p-1}{p \left(e^{8 \gamma t}-1\right)}\right)+e^{-4 \gamma t} \sqrt{(1-2 p)^2 e^{8 \gamma t}-4 (p-1) p} \tanh ^{-1}\left(e^{-4 \gamma t} \sqrt{(1-2 p)^2 e^{8 \gamma t}-4 (p-1) p}\right)\nonumber\\
 &-4 \gamma t+2 p \tanh ^{-1}(1-2 p).
    \end{align}
 For $p=1/2$, we have 
    \begin{align}
         \norm{\ln \rho_t-\ln \sigma_t}_{\rm 2} = \sqrt{\log ^2\left(\frac{1}{2} \left(e^{-4 \gamma t}+1\right)\right)+\log ^2\left(\frac{1}{2}-\frac{1}{2} e^{-4 \gamma t}\right)+\log (4) \log \left(\frac{1}{4}-\frac{1}{4} e^{-8 \gamma t}\right)+2 \log ^2(2)}.  \end{align}
  For $p=0.3$, we have 
         \begin{align}
        (\norm{\ln \rho_t-\ln \sigma_t}_{\rm 2})^2 &=\log \left(\frac{1}{2}-\frac{1}{10} \sqrt{y} e^{-4 \gamma t}\right)^2+\log \left(\frac{1}{10} \left(\sqrt{y} e^{-4 \gamma t}+5\right)\right) \left(\log \left(\frac{10}{21} \left(\sqrt{y} e^{-4 \gamma t}+5\right)\right)-\frac{2 \log \left(\frac{7}{3}\right) e^{4 \gamma t}}{\sqrt{y}}\right)\nonumber\\
        &+\log \left(\frac{1}{2}-\frac{1}{10} \sqrt{y} e^{-4 \gamma t}\right) \left(\frac{\log \left(\frac{49}{9}\right) e^{4 \gamma t}}{\sqrt{y}}+\log \left(\frac{100}{21}\right)\right)+\log \left(\frac{10}{3}\right)^2+\log \left(\frac{10}{7}\right)^2, \end{align}
        where we have defined $y:=4 e^{8 \gamma t}+21$. For $p=0.4$ we have 
        \begin{align}
        (\norm{\ln \rho_t-\ln \sigma_t}_{\rm 2})^2 &= \log ^2\left(\frac{1}{2}-\frac{1}{10} \sqrt{z} e^{-4 \gamma t}\right)+\log \left(\frac{1}{10} \left(\sqrt{z} e^{-4 \gamma t}+5\right)\right) \left(\log \left(\frac{5}{12} \left(\sqrt{z} e^{-4 \gamma t}+5\right)\right)-\frac{\log \left(\frac{3}{2}\right) e^{4 \gamma t}}{\sqrt{z}}\right)\nonumber\\
        &+\left(\frac{\log \left(\frac{3}{2}\right) e^{4 \gamma t}}{\sqrt{z}}+\log \left(\frac{25}{6}\right)\right) \log \left(\frac{1}{2}-\frac{1}{10} \sqrt{z} e^{-4 \gamma t}\right)+\log ^2\left(\frac{5}{2}\right)+\log ^2\left(\frac{5}{3}\right),\end{align}
        where $z:=e^{8 \gamma t}+24$.
 
\twocolumngrid

\bibliography{name.bib}

\begin{thebibliography}{100}%
\makeatletter
\providecommand \@ifxundefined [1]{%
 \@ifx{#1\undefined}
}%
\providecommand \@ifnum [1]{%
 \ifnum #1\expandafter \@firstoftwo
 \else \expandafter \@secondoftwo
 \fi
}%
\providecommand \@ifx [1]{%
 \ifx #1\expandafter \@firstoftwo
 \else \expandafter \@secondoftwo
 \fi
}%
\providecommand \natexlab [1]{#1}%
\providecommand \enquote  [1]{``#1''}%
\providecommand \bibnamefont  [1]{#1}%
\providecommand \bibfnamefont [1]{#1}%
\providecommand \citenamefont [1]{#1}%
\providecommand \href@noop [0]{\@secondoftwo}%
\providecommand \href [0]{\begingroup \@sanitize@url \@href}%
\providecommand \@href[1]{\@@startlink{#1}\@@href}%
\providecommand \@@href[1]{\endgroup#1\@@endlink}%
\providecommand \@sanitize@url [0]{\catcode `\\12\catcode `\$12\catcode
  `\&12\catcode `\#12\catcode `\^12\catcode `\_12\catcode `\%12\relax}%
\providecommand \@@startlink[1]{}%
\providecommand \@@endlink[0]{}%
\providecommand \url  [0]{\begingroup\@sanitize@url \@url }%
\providecommand \@url [1]{\endgroup\@href {#1}{\urlprefix }}%
\providecommand \urlprefix  [0]{URL }%
\providecommand \Eprint [0]{\href }%
\providecommand \doibase [0]{https://doi.org/}%
\providecommand \selectlanguage [0]{\@gobble}%
\providecommand \bibinfo  [0]{\@secondoftwo}%
\providecommand \bibfield  [0]{\@secondoftwo}%
\providecommand \translation [1]{[#1]}%
\providecommand \BibitemOpen [0]{}%
\providecommand \bibitemStop [0]{}%
\providecommand \bibitemNoStop [0]{.\EOS\space}%
\providecommand \EOS [0]{\spacefactor3000\relax}%
\providecommand \BibitemShut  [1]{\csname bibitem#1\endcsname}%
\let\auto@bib@innerbib\@empty
\bibitem [{\citenamefont {Mandelstam}\ and\ \citenamefont
  {Tamm}(1945)}]{Mandelstam1945}%
  \BibitemOpen
  \bibfield  {author} {\bibinfo {author} {\bibfnamefont {L.}~\bibnamefont
  {Mandelstam}}\ and\ \bibinfo {author} {\bibfnamefont {I.}~\bibnamefont
  {Tamm}},\ }\bibfield  {title} {\bibinfo {title} {The uncertainty relation
  between energy and time in non-relativistic quantum mechanics},\ }\href
  {https://doi.org/10.1007/978-3-642-74626-0_8} {\bibfield  {journal} {\bibinfo
   {journal} {J. Phys. (USSR)}\ }\textbf {\bibinfo {volume} {9}},\ \bibinfo
  {pages} {249} (\bibinfo {year} {1945})}\BibitemShut {NoStop}%
\bibitem [{\citenamefont {Margolus}\ and\ \citenamefont
  {Levitin}(1998)}]{Margolus1998}%
  \BibitemOpen
  \bibfield  {author} {\bibinfo {author} {\bibfnamefont {N.}~\bibnamefont
  {Margolus}}\ and\ \bibinfo {author} {\bibfnamefont {L.~B.}\ \bibnamefont
  {Levitin}},\ }\bibfield  {title} {\bibinfo {title} {The maximum speed of
  dynamical evolution},\ }\href
  {https://doi.org/https://doi.org/10.1016/S0167-2789(98)00054-2} {\bibfield
  {journal} {\bibinfo  {journal} {Physica D: Nonlinear Phenomena}\ }\textbf
  {\bibinfo {volume} {120}},\ \bibinfo {pages} {188} (\bibinfo {year}
  {1998})}\BibitemShut {NoStop}%
\bibitem [{\citenamefont {Anandan}\ and\ \citenamefont
  {Aharonov}(1990)}]{Anandan1990}%
  \BibitemOpen
  \bibfield  {author} {\bibinfo {author} {\bibfnamefont {J.}~\bibnamefont
  {Anandan}}\ and\ \bibinfo {author} {\bibfnamefont {Y.}~\bibnamefont
  {Aharonov}},\ }\bibfield  {title} {\bibinfo {title} {Geometry of quantum
  evolution},\ }\href {https://doi.org/10.1103/PhysRevLett.65.1697} {\bibfield
  {journal} {\bibinfo  {journal} {Physical Review Letters}\ }\textbf {\bibinfo
  {volume} {65}},\ \bibinfo {pages} {1697} (\bibinfo {year}
  {1990})}\BibitemShut {NoStop}%
\bibitem [{\citenamefont {Pati}\ \emph {et~al.}(2023)\citenamefont {Pati},
  \citenamefont {Mohan}, \citenamefont {Sahil},\ and\ \citenamefont
  {Braunstein}}]{Pati2023}%
  \BibitemOpen
  \bibfield  {author} {\bibinfo {author} {\bibfnamefont {A.~K.}\ \bibnamefont
  {Pati}}, \bibinfo {author} {\bibfnamefont {B.}~\bibnamefont {Mohan}},
  \bibinfo {author} {\bibnamefont {Sahil}},\ and\ \bibinfo {author}
  {\bibfnamefont {S.~L.}\ \bibnamefont {Braunstein}},\ }\bibfield  {title}
  {\bibinfo {title} {Stronger quantum speed limit},\ }\href
  {https://arxiv.org/abs/2305.03839} {\bibfield  {journal} {\bibinfo  {journal}
  {arXiv:2305.03839}\ } (\bibinfo {year} {2023})}\BibitemShut {NoStop}%
\bibitem [{\citenamefont {Thakuria}\ and\ \citenamefont
  {Pati}(2022)}]{Dimpi20222}%
  \BibitemOpen
  \bibfield  {author} {\bibinfo {author} {\bibfnamefont {D.}~\bibnamefont
  {Thakuria}}\ and\ \bibinfo {author} {\bibfnamefont {A.~K.}\ \bibnamefont
  {Pati}},\ }\bibfield  {title} {\bibinfo {title} {Stronger quantum speed
  limit},\ }\href {https://doi.org/10.48550/arXiv.2208.05469} {\bibfield
  {journal} {\bibinfo  {journal} {arXiv:2208.05469}\ } (\bibinfo {year}
  {2022})}\BibitemShut {NoStop}%
\bibitem [{\citenamefont {Uhlmann}(1992)}]{UHLMANN1992}%
  \BibitemOpen
  \bibfield  {author} {\bibinfo {author} {\bibfnamefont {A.}~\bibnamefont
  {Uhlmann}},\ }\bibfield  {title} {\bibinfo {title} {An energy dispersion
  estimate},\ }\href
  {https://doi.org/https://doi.org/10.1016/0375-9601(92)90555-Z} {\bibfield
  {journal} {\bibinfo  {journal} {Physics Letters A}\ }\textbf {\bibinfo
  {volume} {161}},\ \bibinfo {pages} {329} (\bibinfo {year}
  {1992})}\BibitemShut {NoStop}%
\bibitem [{\citenamefont {Giovannetti}\ \emph {et~al.}(2004)\citenamefont
  {Giovannetti}, \citenamefont {Lloyd},\ and\ \citenamefont
  {Maccone}}]{Giovannetti_2004}%
  \BibitemOpen
  \bibfield  {author} {\bibinfo {author} {\bibfnamefont {V.}~\bibnamefont
  {Giovannetti}}, \bibinfo {author} {\bibfnamefont {S.}~\bibnamefont {Lloyd}},\
  and\ \bibinfo {author} {\bibfnamefont {L.}~\bibnamefont {Maccone}},\
  }\bibfield  {title} {\bibinfo {title} {The speed limit of quantum unitary
  evolution},\ }\href {https://doi.org/10.1088/1464-4266/6/8/028} {\bibfield
  {journal} {\bibinfo  {journal} {Journal of Optics B: Quantum and
  Semiclassical Optics}\ }\textbf {\bibinfo {volume} {6}},\ \bibinfo {pages}
  {S807} (\bibinfo {year} {2004})}\BibitemShut {NoStop}%
\bibitem [{\citenamefont {J}\ and\ \citenamefont
  {Hern{\'{a}}ndez}(2022)}]{Canseco_J_2022}%
  \BibitemOpen
  \bibfield  {author} {\bibinfo {author} {\bibfnamefont {S.~C.}\ \bibnamefont
  {J}}\ and\ \bibinfo {author} {\bibfnamefont {A.~V.}\ \bibnamefont
  {Hern{\'{a}}ndez}},\ }\bibfield  {title} {\bibinfo {title} {Speed of
  evolution in entangled fermionic systems},\ }\href
  {https://doi.org/10.1088/1751-8121/ac8ef8} {\bibfield  {journal} {\bibinfo
  {journal} {Journal of Physics A: Mathematical and Theoretical}\ }\textbf
  {\bibinfo {volume} {55}},\ \bibinfo {pages} {405301} (\bibinfo {year}
  {2022})}\BibitemShut {NoStop}%
\bibitem [{\citenamefont {Krisnanda}\ \emph {et~al.}(2022)\citenamefont
  {Krisnanda}, \citenamefont {Lee}, \citenamefont {Noh}, \citenamefont {Kim},
  \citenamefont {Streltsov}, \citenamefont {Liew},\ and\ \citenamefont
  {Paterek}}]{Krisnanda_2022}%
  \BibitemOpen
  \bibfield  {author} {\bibinfo {author} {\bibfnamefont {T.}~\bibnamefont
  {Krisnanda}}, \bibinfo {author} {\bibfnamefont {S.}~\bibnamefont {Lee}},
  \bibinfo {author} {\bibfnamefont {C.}~\bibnamefont {Noh}}, \bibinfo {author}
  {\bibfnamefont {J.}~\bibnamefont {Kim}}, \bibinfo {author} {\bibfnamefont
  {A.}~\bibnamefont {Streltsov}}, \bibinfo {author} {\bibfnamefont {T.~C.~H.}\
  \bibnamefont {Liew}},\ and\ \bibinfo {author} {\bibfnamefont
  {T.}~\bibnamefont {Paterek}},\ }\bibfield  {title} {\bibinfo {title}
  {Correlations and energy in mediated dynamics},\ }\href
  {https://doi.org/10.1088/1367-2630/aca9ef} {\bibfield  {journal} {\bibinfo
  {journal} {New Journal of Physics}\ }\textbf {\bibinfo {volume} {24}},\
  \bibinfo {pages} {123025} (\bibinfo {year} {2022})}\BibitemShut {NoStop}%
\bibitem [{\citenamefont {Taddei}\ \emph {et~al.}(2013)\citenamefont {Taddei},
  \citenamefont {Escher}, \citenamefont {Davidovich},\ and\ \citenamefont
  {de~Matos~Filho}}]{Taddei2013}%
  \BibitemOpen
  \bibfield  {author} {\bibinfo {author} {\bibfnamefont {M.~M.}\ \bibnamefont
  {Taddei}}, \bibinfo {author} {\bibfnamefont {B.~M.}\ \bibnamefont {Escher}},
  \bibinfo {author} {\bibfnamefont {L.}~\bibnamefont {Davidovich}},\ and\
  \bibinfo {author} {\bibfnamefont {R.~L.}\ \bibnamefont {de~Matos~Filho}},\
  }\bibfield  {title} {\bibinfo {title} {Quantum speed limit for physical
  processes},\ }\href {https://doi.org/10.1103/PhysRevLett.110.050402}
  {\bibfield  {journal} {\bibinfo  {journal} {Physical Review Letters}\
  }\textbf {\bibinfo {volume} {110}},\ \bibinfo {pages} {050402} (\bibinfo
  {year} {2013})}\BibitemShut {NoStop}%
\bibitem [{\citenamefont {del Campo}\ \emph {et~al.}(2013)\citenamefont {del
  Campo}, \citenamefont {Egusquiza}, \citenamefont {Plenio},\ and\
  \citenamefont {Huelga}}]{del_Campo2013}%
  \BibitemOpen
  \bibfield  {author} {\bibinfo {author} {\bibfnamefont {A.}~\bibnamefont {del
  Campo}}, \bibinfo {author} {\bibfnamefont {I.~L.}\ \bibnamefont {Egusquiza}},
  \bibinfo {author} {\bibfnamefont {M.~B.}\ \bibnamefont {Plenio}},\ and\
  \bibinfo {author} {\bibfnamefont {S.~F.}\ \bibnamefont {Huelga}},\ }\bibfield
   {title} {\bibinfo {title} {Quantum speed limits in open system dynamics},\
  }\href {https://doi.org/10.1103/PhysRevLett.110.050403} {\bibfield  {journal}
  {\bibinfo  {journal} {Physical Review Letters}\ }\textbf {\bibinfo {volume}
  {110}},\ \bibinfo {pages} {050403} (\bibinfo {year} {2013})}\BibitemShut
  {NoStop}%
\bibitem [{\citenamefont {Deffner}\ and\ \citenamefont
  {Lutz}(2013)}]{Deffner_2013}%
  \BibitemOpen
  \bibfield  {author} {\bibinfo {author} {\bibfnamefont {S.}~\bibnamefont
  {Deffner}}\ and\ \bibinfo {author} {\bibfnamefont {E.}~\bibnamefont {Lutz}},\
  }\bibfield  {title} {\bibinfo {title} {Quantum speed limit for non-markovian
  dynamics},\ }\href {https://doi.org/10.1103/PhysRevLett.111.010402}
  {\bibfield  {journal} {\bibinfo  {journal} {Physical Review Letters}\
  }\textbf {\bibinfo {volume} {111}},\ \bibinfo {pages} {010402} (\bibinfo
  {year} {2013})}\BibitemShut {NoStop}%
\bibitem [{\citenamefont {Pires}\ \emph {et~al.}(2016)\citenamefont {Pires},
  \citenamefont {Cianciaruso}, \citenamefont {C\'eleri}, \citenamefont
  {Adesso},\ and\ \citenamefont {Soares-Pinto}}]{Peris2016}%
  \BibitemOpen
  \bibfield  {author} {\bibinfo {author} {\bibfnamefont {D.~P.}\ \bibnamefont
  {Pires}}, \bibinfo {author} {\bibfnamefont {M.}~\bibnamefont {Cianciaruso}},
  \bibinfo {author} {\bibfnamefont {L.~C.}\ \bibnamefont {C\'eleri}}, \bibinfo
  {author} {\bibfnamefont {G.}~\bibnamefont {Adesso}},\ and\ \bibinfo {author}
  {\bibfnamefont {D.~O.}\ \bibnamefont {Soares-Pinto}},\ }\bibfield  {title}
  {\bibinfo {title} {Generalized geometric quantum speed limits},\ }\href
  {https://doi.org/10.1103/PhysRevX.6.021031} {\bibfield  {journal} {\bibinfo
  {journal} {Physical Review X}\ }\textbf {\bibinfo {volume} {6}},\ \bibinfo
  {pages} {021031} (\bibinfo {year} {2016})}\BibitemShut {NoStop}%
\bibitem [{\citenamefont {Shiraishi}\ and\ \citenamefont
  {Saito}(2021)}]{Naoto2021}%
  \BibitemOpen
  \bibfield  {author} {\bibinfo {author} {\bibfnamefont {N.}~\bibnamefont
  {Shiraishi}}\ and\ \bibinfo {author} {\bibfnamefont {K.}~\bibnamefont
  {Saito}},\ }\bibfield  {title} {\bibinfo {title} {Speed limit for open
  systems coupled to general environments},\ }\href
  {https://doi.org/10.1103/PhysRevResearch.3.023074} {\bibfield  {journal}
  {\bibinfo  {journal} {Physical Review Research}\ }\textbf {\bibinfo {volume}
  {3}},\ \bibinfo {pages} {023074} (\bibinfo {year} {2021})}\BibitemShut
  {NoStop}%
\bibitem [{\citenamefont {Funo}\ \emph {et~al.}(2019)\citenamefont {Funo},
  \citenamefont {Shiraishi},\ and\ \citenamefont {Saito}}]{Funo_2019}%
  \BibitemOpen
  \bibfield  {author} {\bibinfo {author} {\bibfnamefont {K.}~\bibnamefont
  {Funo}}, \bibinfo {author} {\bibfnamefont {N.}~\bibnamefont {Shiraishi}},\
  and\ \bibinfo {author} {\bibfnamefont {K.}~\bibnamefont {Saito}},\ }\bibfield
   {title} {\bibinfo {title} {Speed limit for open quantum systems},\ }\href
  {https://doi.org/10.1088/1367-2630/aaf9f5} {\bibfield  {journal} {\bibinfo
  {journal} {New Journal of Physics}\ }\textbf {\bibinfo {volume} {21}},\
  \bibinfo {pages} {013006} (\bibinfo {year} {2019})}\BibitemShut {NoStop}%
\bibitem [{\citenamefont {Van~Vu}\ and\ \citenamefont
  {Saito}(2023)}]{Saito2023}%
  \BibitemOpen
  \bibfield  {author} {\bibinfo {author} {\bibfnamefont {T.}~\bibnamefont
  {Van~Vu}}\ and\ \bibinfo {author} {\bibfnamefont {K.}~\bibnamefont {Saito}},\
  }\bibfield  {title} {\bibinfo {title} {Topological speed limit},\ }\href
  {https://doi.org/10.1103/PhysRevLett.130.010402} {\bibfield  {journal}
  {\bibinfo  {journal} {Physical Review Letters}\ }\textbf {\bibinfo {volume}
  {130}},\ \bibinfo {pages} {010402} (\bibinfo {year} {2023})}\BibitemShut
  {NoStop}%
\bibitem [{\citenamefont {Brody}\ and\ \citenamefont
  {Longstaff}(2019)}]{Brody_2019}%
  \BibitemOpen
  \bibfield  {author} {\bibinfo {author} {\bibfnamefont {D.~C.}\ \bibnamefont
  {Brody}}\ and\ \bibinfo {author} {\bibfnamefont {B.}~\bibnamefont
  {Longstaff}},\ }\bibfield  {title} {\bibinfo {title} {Evolution speed of open
  quantum dynamics},\ }\bibfield  {journal} {\bibinfo  {journal} {Physical
  Review Research}\ }\textbf {\bibinfo {volume} {1}},\ \href
  {https://doi.org/10.1103/physrevresearch.1.033127}
  {10.1103/physrevresearch.1.033127} (\bibinfo {year} {2019})\BibitemShut
  {NoStop}%
\bibitem [{\citenamefont {Campaioli}\ \emph {et~al.}(2013)\citenamefont
  {Campaioli}, \citenamefont {Pollock},\ and\ \citenamefont
  {Modi}}]{Campaioli2013}%
  \BibitemOpen
  \bibfield  {author} {\bibinfo {author} {\bibfnamefont {F.}~\bibnamefont
  {Campaioli}}, \bibinfo {author} {\bibfnamefont {F.}~\bibnamefont {Pollock}},\
  and\ \bibinfo {author} {\bibfnamefont {K.}~\bibnamefont {Modi}},\ }\bibfield
  {title} {\bibinfo {title} {Quantum speed limits in open system dynamics},\
  }\href {https://doi.org/https://doi.org/10.22331/q-2019-08-05-168} {\bibfield
   {journal} {\bibinfo  {journal} {Quantum}\ }\textbf {\bibinfo {volume} {3}},\
  \bibinfo {pages} {168} (\bibinfo {year} {2013})}\BibitemShut {NoStop}%
\bibitem [{\citenamefont {Kiselev}\ \emph {et~al.}(2022)\citenamefont
  {Kiselev}, \citenamefont {R.},\ and\ \citenamefont {Rybin}}]{Kiselev2022}%
  \BibitemOpen
  \bibfield  {author} {\bibinfo {author} {\bibfnamefont {A.~D.}\ \bibnamefont
  {Kiselev}}, \bibinfo {author} {\bibfnamefont {A.}~\bibnamefont {R.}},\ and\
  \bibinfo {author} {\bibfnamefont {A.~V.}\ \bibnamefont {Rybin}},\ }\bibfield
  {title} {\bibinfo {title} {Speed of evolution and correlations in multi-mode
  bosonic systems},\ }\href {https://www.mdpi.com/1099-4300/24/12/1774}
  {\bibfield  {journal} {\bibinfo  {journal} {Entropy,}\ }\textbf {\bibinfo
  {volume} {24}} (\bibinfo {year} {2022})}\BibitemShut {NoStop}%
\bibitem [{\citenamefont {Thakuria}\ \emph {et~al.}(2022)\citenamefont
  {Thakuria}, \citenamefont {Srivastav}, \citenamefont {Mohan}, \citenamefont
  {Kumari},\ and\ \citenamefont {Pati}}]{Dimpi2022}%
  \BibitemOpen
  \bibfield  {author} {\bibinfo {author} {\bibfnamefont {D.}~\bibnamefont
  {Thakuria}}, \bibinfo {author} {\bibfnamefont {A.}~\bibnamefont {Srivastav}},
  \bibinfo {author} {\bibfnamefont {B.}~\bibnamefont {Mohan}}, \bibinfo
  {author} {\bibfnamefont {A.}~\bibnamefont {Kumari}},\ and\ \bibinfo {author}
  {\bibfnamefont {A.~K.}\ \bibnamefont {Pati}},\ }\bibfield  {title} {\bibinfo
  {title} {Generalised quantum speed limit for arbitrary evolution},\ }\href
  {https://doi.org/10.48550/arXiv.2207.04124} {\bibfield  {journal} {\bibinfo
  {journal} {arXiv:2207.04124}\ } (\bibinfo {year} {2022})}\BibitemShut
  {NoStop}%
\bibitem [{\citenamefont {Ashhab}\ \emph {et~al.}(2012)\citenamefont {Ashhab},
  \citenamefont {de~Groot},\ and\ \citenamefont {Nori}}]{AGN12}%
  \BibitemOpen
  \bibfield  {author} {\bibinfo {author} {\bibfnamefont {S.}~\bibnamefont
  {Ashhab}}, \bibinfo {author} {\bibfnamefont {P.~C.}\ \bibnamefont
  {de~Groot}},\ and\ \bibinfo {author} {\bibfnamefont {F.}~\bibnamefont
  {Nori}},\ }\bibfield  {title} {\bibinfo {title} {Speed limits for quantum
  gates in multiqubit systems},\ }\href
  {https://doi.org/10.1103/PhysRevA.85.052327} {\bibfield  {journal} {\bibinfo
  {journal} {Physical Review A}\ }\textbf {\bibinfo {volume} {85}},\ \bibinfo
  {pages} {052327} (\bibinfo {year} {2012})}\BibitemShut {NoStop}%
\bibitem [{\citenamefont {Aifer}\ and\ \citenamefont
  {Deffner}(2022)}]{Aifer2022}%
  \BibitemOpen
  \bibfield  {author} {\bibinfo {author} {\bibfnamefont {M.}~\bibnamefont
  {Aifer}}\ and\ \bibinfo {author} {\bibfnamefont {S.}~\bibnamefont
  {Deffner}},\ }\bibfield  {title} {\bibinfo {title} {From quantum speed limits
  to energy-efficient quantum gates},\ }\href
  {https://doi.org/10.1088/1367-2630/ac6821} {\bibfield  {journal} {\bibinfo
  {journal} {New Journal of Physics}\ }\textbf {\bibinfo {volume} {24}},\
  \bibinfo {pages} {055002} (\bibinfo {year} {2022})}\BibitemShut {NoStop}%
\bibitem [{\citenamefont {Campbell}\ \emph {et~al.}(2018)\citenamefont
  {Campbell}, \citenamefont {Genoni},\ and\ \citenamefont
  {Deffner}}]{Campbell2018}%
  \BibitemOpen
  \bibfield  {author} {\bibinfo {author} {\bibfnamefont {S.}~\bibnamefont
  {Campbell}}, \bibinfo {author} {\bibfnamefont {M.~G.}\ \bibnamefont
  {Genoni}},\ and\ \bibinfo {author} {\bibfnamefont {S.}~\bibnamefont
  {Deffner}},\ }\bibfield  {title} {\bibinfo {title} {Precision thermometry and
  the quantum speed limit},\ }\href {https://doi.org/10.1088/2058-9565/aaa641}
  {\bibfield  {journal} {\bibinfo  {journal} {Quantum Science and Technology}\
  }\textbf {\bibinfo {volume} {3}},\ \bibinfo {pages} {025002} (\bibinfo {year}
  {2018})}\BibitemShut {NoStop}%
\bibitem [{\citenamefont {Beau}\ and\ \citenamefont {del
  Campo}(2017)}]{Beau2017}%
  \BibitemOpen
  \bibfield  {author} {\bibinfo {author} {\bibfnamefont {M.}~\bibnamefont
  {Beau}}\ and\ \bibinfo {author} {\bibfnamefont {A.}~\bibnamefont {del
  Campo}},\ }\bibfield  {title} {\bibinfo {title} {Nonlinear quantum metrology
  of many-body open systems},\ }\href
  {https://doi.org/10.1103/PhysRevLett.119.010403} {\bibfield  {journal}
  {\bibinfo  {journal} {Physical Review Letters}\ }\textbf {\bibinfo {volume}
  {119}},\ \bibinfo {pages} {010403} (\bibinfo {year} {2017})}\BibitemShut
  {NoStop}%
\bibitem [{\citenamefont {Caneva}\ \emph {et~al.}(2009)\citenamefont {Caneva},
  \citenamefont {Murphy}, \citenamefont {Calarco}, \citenamefont {Fazio},
  \citenamefont {Montangero}, \citenamefont {Giovannetti},\ and\ \citenamefont
  {Santoro}}]{Caneva2009}%
  \BibitemOpen
  \bibfield  {author} {\bibinfo {author} {\bibfnamefont {T.}~\bibnamefont
  {Caneva}}, \bibinfo {author} {\bibfnamefont {M.}~\bibnamefont {Murphy}},
  \bibinfo {author} {\bibfnamefont {T.}~\bibnamefont {Calarco}}, \bibinfo
  {author} {\bibfnamefont {R.}~\bibnamefont {Fazio}}, \bibinfo {author}
  {\bibfnamefont {S.}~\bibnamefont {Montangero}}, \bibinfo {author}
  {\bibfnamefont {V.}~\bibnamefont {Giovannetti}},\ and\ \bibinfo {author}
  {\bibfnamefont {G.~E.}\ \bibnamefont {Santoro}},\ }\bibfield  {title}
  {\bibinfo {title} {Optimal control at the quantum speed limit},\ }\href
  {https://doi.org/10.1103/PhysRevLett.103.240501} {\bibfield  {journal}
  {\bibinfo  {journal} {Physical Review Letters}\ }\textbf {\bibinfo {volume}
  {103}},\ \bibinfo {pages} {240501} (\bibinfo {year} {2009})}\BibitemShut
  {NoStop}%
\bibitem [{\citenamefont {Deffner}\ and\ \citenamefont
  {Campbell}(2017)}]{deffner_2017}%
  \BibitemOpen
  \bibfield  {author} {\bibinfo {author} {\bibfnamefont {S.}~\bibnamefont
  {Deffner}}\ and\ \bibinfo {author} {\bibfnamefont {S.}~\bibnamefont
  {Campbell}},\ }\bibfield  {title} {\bibinfo {title} {Quantum speed limits:
  from heisenberg's uncertainty principle to optimal quantum control},\ }\href
  {https://doi.org/10.1088/1751-8121/aa86c6} {\bibfield  {journal} {\bibinfo
  {journal} {Journal of Physics A: Mathematical and Theoretical}\ }\textbf
  {\bibinfo {volume} {50}},\ \bibinfo {pages} {453001} (\bibinfo {year}
  {2017})}\BibitemShut {NoStop}%
\bibitem [{\citenamefont {Murphy}\ \emph {et~al.}(2010)\citenamefont {Murphy},
  \citenamefont {Montangero}, \citenamefont {Giovannetti},\ and\ \citenamefont
  {Calarco}}]{Murphy2010}%
  \BibitemOpen
  \bibfield  {author} {\bibinfo {author} {\bibfnamefont {M.}~\bibnamefont
  {Murphy}}, \bibinfo {author} {\bibfnamefont {S.}~\bibnamefont {Montangero}},
  \bibinfo {author} {\bibfnamefont {V.}~\bibnamefont {Giovannetti}},\ and\
  \bibinfo {author} {\bibfnamefont {T.}~\bibnamefont {Calarco}},\ }\bibfield
  {title} {\bibinfo {title} {Communication at the quantum speed limit along a
  spin chain},\ }\href {https://doi.org/10.1103/PhysRevA.82.022318} {\bibfield
  {journal} {\bibinfo  {journal} {Physical Review A}\ }\textbf {\bibinfo
  {volume} {82}},\ \bibinfo {pages} {022318} (\bibinfo {year}
  {2010})}\BibitemShut {NoStop}%
\bibitem [{\citenamefont {Mohan}\ and\ \citenamefont
  {Pati}(2021)}]{Mohan_Pati_2021}%
  \BibitemOpen
  \bibfield  {author} {\bibinfo {author} {\bibfnamefont {B.}~\bibnamefont
  {Mohan}}\ and\ \bibinfo {author} {\bibfnamefont {A.~K.}\ \bibnamefont
  {Pati}},\ }\bibfield  {title} {\bibinfo {title} {Reverse quantum speed limit:
  How slowly a quantum battery can discharge},\ }\href
  {https://doi.org/10.1103/PhysRevA.104.042209} {\bibfield  {journal} {\bibinfo
   {journal} {Physical Review A}\ }\textbf {\bibinfo {volume} {104}},\ \bibinfo
  {pages} {042209} (\bibinfo {year} {2021})}\BibitemShut {NoStop}%
\bibitem [{\citenamefont {Mohan}\ and\ \citenamefont {Pati}(2022)}]{Mohan2022}%
  \BibitemOpen
  \bibfield  {author} {\bibinfo {author} {\bibfnamefont {B.}~\bibnamefont
  {Mohan}}\ and\ \bibinfo {author} {\bibfnamefont {A.~K.}\ \bibnamefont
  {Pati}},\ }\bibfield  {title} {\bibinfo {title} {Quantum speed limits for
  observables},\ }\href {https://doi.org/10.1103/PhysRevA.106.042436}
  {\bibfield  {journal} {\bibinfo  {journal} {Physical Review A}\ }\textbf
  {\bibinfo {volume} {106}},\ \bibinfo {pages} {042436} (\bibinfo {year}
  {2022})}\BibitemShut {NoStop}%
\bibitem [{\citenamefont {Campaioli}\ \emph {et~al.}(2017)\citenamefont
  {Campaioli}, \citenamefont {Pollock}, \citenamefont {Binder}, \citenamefont
  {C\'eleri}, \citenamefont {Goold}, \citenamefont {Vinjanampathy},\ and\
  \citenamefont {Modi}}]{Modi2017}%
  \BibitemOpen
  \bibfield  {author} {\bibinfo {author} {\bibfnamefont {F.}~\bibnamefont
  {Campaioli}}, \bibinfo {author} {\bibfnamefont {F.~A.}\ \bibnamefont
  {Pollock}}, \bibinfo {author} {\bibfnamefont {F.~C.}\ \bibnamefont {Binder}},
  \bibinfo {author} {\bibfnamefont {L.}~\bibnamefont {C\'eleri}}, \bibinfo
  {author} {\bibfnamefont {J.}~\bibnamefont {Goold}}, \bibinfo {author}
  {\bibfnamefont {S.}~\bibnamefont {Vinjanampathy}},\ and\ \bibinfo {author}
  {\bibfnamefont {K.}~\bibnamefont {Modi}},\ }\bibfield  {title} {\bibinfo
  {title} {Enhancing the charging power of quantum batteries},\ }\href
  {https://doi.org/10.1103/PhysRevLett.118.150601} {\bibfield  {journal}
  {\bibinfo  {journal} {Physical Review Letters}\ }\textbf {\bibinfo {volume}
  {118}},\ \bibinfo {pages} {150601} (\bibinfo {year} {2017})}\BibitemShut
  {NoStop}%
\bibitem [{\citenamefont {Mukhopadhyay}\ \emph {et~al.}(2018)\citenamefont
  {Mukhopadhyay}, \citenamefont {Misra}, \citenamefont {Bhattacharya},\ and\
  \citenamefont {Pati}}]{Mukhopadhyay2018}%
  \BibitemOpen
  \bibfield  {author} {\bibinfo {author} {\bibfnamefont {C.}~\bibnamefont
  {Mukhopadhyay}}, \bibinfo {author} {\bibfnamefont {A.}~\bibnamefont {Misra}},
  \bibinfo {author} {\bibfnamefont {S.}~\bibnamefont {Bhattacharya}},\ and\
  \bibinfo {author} {\bibfnamefont {A.~K.}\ \bibnamefont {Pati}},\ }\bibfield
  {title} {\bibinfo {title} {Quantum speed limit constraints on a nanoscale
  autonomous refrigerator},\ }\href
  {https://doi.org/10.1103/PhysRevE.97.062116} {\bibfield  {journal} {\bibinfo
  {journal} {Physical Review E}\ }\textbf {\bibinfo {volume} {97}},\ \bibinfo
  {pages} {062116} (\bibinfo {year} {2018})}\BibitemShut {NoStop}%
\bibitem [{\citenamefont {Einstein}\ \emph {et~al.}(1935)\citenamefont
  {Einstein}, \citenamefont {Podolsky},\ and\ \citenamefont {Rosen}}]{EPR1935}%
  \BibitemOpen
  \bibfield  {author} {\bibinfo {author} {\bibfnamefont {A.}~\bibnamefont
  {Einstein}}, \bibinfo {author} {\bibfnamefont {B.}~\bibnamefont {Podolsky}},\
  and\ \bibinfo {author} {\bibfnamefont {N.}~\bibnamefont {Rosen}},\ }\bibfield
   {title} {\bibinfo {title} {Can quantum-mechanical description of physical
  reality be considered complete?},\ }\href
  {https://doi.org/10.1103/PhysRev.47.777} {\bibfield  {journal} {\bibinfo
  {journal} {Physical Review}\ }\textbf {\bibinfo {volume} {47}},\ \bibinfo
  {pages} {777} (\bibinfo {year} {1935})}\BibitemShut {NoStop}%
\bibitem [{\citenamefont {Bell}(1964)}]{Bell1964}%
  \BibitemOpen
  \bibfield  {author} {\bibinfo {author} {\bibfnamefont {J.~S.}\ \bibnamefont
  {Bell}},\ }\bibfield  {title} {\bibinfo {title} {On the {E}instein {P}odolsky
  {R}osen paradox},\ }\href
  {https://doi.org/10.1103/PhysicsPhysiqueFizika.1.195} {\bibfield  {journal}
  {\bibinfo  {journal} {Physics Physique Fizika}\ }\textbf {\bibinfo {volume}
  {1}},\ \bibinfo {pages} {195} (\bibinfo {year} {1964})}\BibitemShut {NoStop}%
\bibitem [{\citenamefont {Bennett}\ \emph {et~al.}(1993)\citenamefont
  {Bennett}, \citenamefont {Brassard}, \citenamefont {Cr\'epeau}, \citenamefont
  {Jozsa}, \citenamefont {Peres},\ and\ \citenamefont
  {Wootters}}]{Bennett1993a}%
  \BibitemOpen
  \bibfield  {author} {\bibinfo {author} {\bibfnamefont {C.~H.}\ \bibnamefont
  {Bennett}}, \bibinfo {author} {\bibfnamefont {G.}~\bibnamefont {Brassard}},
  \bibinfo {author} {\bibfnamefont {C.}~\bibnamefont {Cr\'epeau}}, \bibinfo
  {author} {\bibfnamefont {R.}~\bibnamefont {Jozsa}}, \bibinfo {author}
  {\bibfnamefont {A.}~\bibnamefont {Peres}},\ and\ \bibinfo {author}
  {\bibfnamefont {W.~K.}\ \bibnamefont {Wootters}},\ }\bibfield  {title}
  {\bibinfo {title} {Teleporting an unknown quantum state via dual classical
  and einstein-podolsky-rosen channels},\ }\href
  {https://doi.org/10.1103/PhysRevLett.70.1895} {\bibfield  {journal} {\bibinfo
   {journal} {Physical Review Letters}\ }\textbf {\bibinfo {volume} {70}},\
  \bibinfo {pages} {1895} (\bibinfo {year} {1993})}\BibitemShut {NoStop}%
\bibitem [{\citenamefont {Bennett}\ and\ \citenamefont
  {Wiesner}(1992)}]{Wiesner1992}%
  \BibitemOpen
  \bibfield  {author} {\bibinfo {author} {\bibfnamefont {C.~H.}\ \bibnamefont
  {Bennett}}\ and\ \bibinfo {author} {\bibfnamefont {S.~J.}\ \bibnamefont
  {Wiesner}},\ }\bibfield  {title} {\bibinfo {title} {Communication via one-
  and two-particle operators on einstein-podolsky-rosen states},\ }\href
  {https://doi.org/10.1103/PhysRevLett.69.2881} {\bibfield  {journal} {\bibinfo
   {journal} {Physical Review Letters}\ }\textbf {\bibinfo {volume} {69}},\
  \bibinfo {pages} {2881} (\bibinfo {year} {1992})}\BibitemShut {NoStop}%
\bibitem [{\citenamefont {Ekert}(1991)}]{Ekert1991}%
  \BibitemOpen
  \bibfield  {author} {\bibinfo {author} {\bibfnamefont {A.~K.}\ \bibnamefont
  {Ekert}},\ }\bibfield  {title} {\bibinfo {title} {Quantum cryptography based
  on bell's theorem},\ }\href {https://doi.org/10.1103/PhysRevLett.67.661}
  {\bibfield  {journal} {\bibinfo  {journal} {Physical Review Letters}\
  }\textbf {\bibinfo {volume} {67}},\ \bibinfo {pages} {661} (\bibinfo {year}
  {1991})}\BibitemShut {NoStop}%
\bibitem [{\citenamefont {Jozsa}(1997)}]{Jozsa1997}%
  \BibitemOpen
  \bibfield  {author} {\bibinfo {author} {\bibfnamefont {R.}~\bibnamefont
  {Jozsa}},\ }\bibfield  {title} {\bibinfo {title} {Entanglement and quantum
  computation},\ }\href {https://doi.org/10.48550/arXiv.quant-ph/9707034}
  {\bibfield  {journal} {\bibinfo  {journal} {quant-ph/9707034}\ } (\bibinfo
  {year} {1997})}\BibitemShut {NoStop}%
\bibitem [{\citenamefont {Jozsa}\ and\ \citenamefont
  {Linden}(2003)}]{Jozsa2003}%
  \BibitemOpen
  \bibfield  {author} {\bibinfo {author} {\bibfnamefont {R.}~\bibnamefont
  {Jozsa}}\ and\ \bibinfo {author} {\bibfnamefont {N.}~\bibnamefont {Linden}},\
  }\bibfield  {title} {\bibinfo {title} {On the role of entanglement in
  quantum-computational speed-up},\ }\href
  {https://doi.org/10.1098/rspa.2002.1097} {\bibfield  {journal} {\bibinfo
  {journal} {Proceedings of the Royal Society of London. Series A:
  Mathematical, Physical and Engineering Sciences}\ }\textbf {\bibinfo {volume}
  {459}},\ \bibinfo {pages} {2011} (\bibinfo {year} {2003})}\BibitemShut
  {NoStop}%
\bibitem [{\citenamefont {Colbeck}\ and\ \citenamefont
  {Renner}(2012)}]{Colbeck2012}%
  \BibitemOpen
  \bibfield  {author} {\bibinfo {author} {\bibfnamefont {R.}~\bibnamefont
  {Colbeck}}\ and\ \bibinfo {author} {\bibfnamefont {R.}~\bibnamefont
  {Renner}},\ }\bibfield  {title} {\bibinfo {title} {Free randomness can be
  amplified},\ }\href {https://doi.org/10.1038/nphys2300} {\bibfield  {journal}
  {\bibinfo  {journal} {Nature Physics}\ }\textbf {\bibinfo {volume} {8}},\
  \bibinfo {pages} {450} (\bibinfo {year} {2012})}\BibitemShut {NoStop}%
\bibitem [{\citenamefont {Dowling}(2008)}]{Jonathon2008}%
  \BibitemOpen
  \bibfield  {author} {\bibinfo {author} {\bibfnamefont {J.~P.}\ \bibnamefont
  {Dowling}},\ }\bibfield  {title} {\bibinfo {title} {Quantum optical metrology
  – the lowdown on high-{N00N} states},\ }\href
  {https://doi.org/10.1080/00107510802091298} {\bibfield  {journal} {\bibinfo
  {journal} {Contemporary Physics}\ }\textbf {\bibinfo {volume} {49}},\
  \bibinfo {pages} {125} (\bibinfo {year} {2008})}\BibitemShut {NoStop}%
\bibitem [{\citenamefont {Ho}\ and\ \citenamefont {Abanin}(2017)}]{mbed}%
  \BibitemOpen
  \bibfield  {author} {\bibinfo {author} {\bibfnamefont {W.~W.}\ \bibnamefont
  {Ho}}\ and\ \bibinfo {author} {\bibfnamefont {D.~A.}\ \bibnamefont
  {Abanin}},\ }\bibfield  {title} {\bibinfo {title} {Entanglement dynamics in
  quantum many-body systems},\ }\href
  {https://doi.org/10.1103/PhysRevB.95.094302} {\bibfield  {journal} {\bibinfo
  {journal} {Physical Review B}\ }\textbf {\bibinfo {volume} {95}},\ \bibinfo
  {pages} {094302} (\bibinfo {year} {2017})}\BibitemShut {NoStop}%
\bibitem [{\citenamefont {Elsayed}\ \emph {et~al.}(2018)\citenamefont
  {Elsayed}, \citenamefont {M{\o}lmer},\ and\ \citenamefont
  {Madsen}}]{Elsayed_2018}%
  \BibitemOpen
  \bibfield  {author} {\bibinfo {author} {\bibfnamefont {T.~A.}\ \bibnamefont
  {Elsayed}}, \bibinfo {author} {\bibfnamefont {K.}~\bibnamefont {M{\o}lmer}},\
  and\ \bibinfo {author} {\bibfnamefont {L.~B.}\ \bibnamefont {Madsen}},\
  }\bibfield  {title} {\bibinfo {title} {Entangled quantum dynamics of
  many-body systems using bohmian trajectories},\ }\bibfield  {journal}
  {\bibinfo  {journal} {Scientific Reports}\ }\textbf {\bibinfo {volume} {8}},\
  \href {https://doi.org/10.1038/s41598-018-30730-0}
  {10.1038/s41598-018-30730-0} (\bibinfo {year} {2018})\BibitemShut {NoStop}%
\bibitem [{\citenamefont {Brand\~ao}\ \emph {et~al.}(2020)\citenamefont
  {Brand\~ao}, \citenamefont {Suassuna}, \citenamefont {Melo},\ and\
  \citenamefont {Guerreiro}}]{oped}%
  \BibitemOpen
  \bibfield  {author} {\bibinfo {author} {\bibfnamefont {I.}~\bibnamefont
  {Brand\~ao}}, \bibinfo {author} {\bibfnamefont {B.}~\bibnamefont {Suassuna}},
  \bibinfo {author} {\bibfnamefont {B.}~\bibnamefont {Melo}},\ and\ \bibinfo
  {author} {\bibfnamefont {T.}~\bibnamefont {Guerreiro}},\ }\bibfield  {title}
  {\bibinfo {title} {Entanglement dynamics in dispersive optomechanics:
  Nonclassicality and revival},\ }\href
  {https://doi.org/10.1103/PhysRevResearch.2.043421} {\bibfield  {journal}
  {\bibinfo  {journal} {Physical Review Research}\ }\textbf {\bibinfo {volume}
  {2}},\ \bibinfo {pages} {043421} (\bibinfo {year} {2020})}\BibitemShut
  {NoStop}%
\bibitem [{\citenamefont {Gong}\ and\ \citenamefont
  {Hamazaki}(2022{\natexlab{a}})}]{Gong_2022}%
  \BibitemOpen
  \bibfield  {author} {\bibinfo {author} {\bibfnamefont {Z.}~\bibnamefont
  {Gong}}\ and\ \bibinfo {author} {\bibfnamefont {R.}~\bibnamefont
  {Hamazaki}},\ }\bibfield  {title} {\bibinfo {title} {Bounds in nonequilibrium
  quantum dynamics},\ }\bibfield  {journal} {\bibinfo  {journal} {International
  Journal of Modern Physics B}\ }\textbf {\bibinfo {volume} {36}},\ \href
  {https://doi.org/10.1142/s0217979222300079} {10.1142/s0217979222300079}
  (\bibinfo {year} {2022}{\natexlab{a}})\BibitemShut {NoStop}%
\bibitem [{\citenamefont {Hamazaki}(2023)}]{hamazaki2023quantum}%
  \BibitemOpen
  \bibfield  {author} {\bibinfo {author} {\bibfnamefont {R.}~\bibnamefont
  {Hamazaki}},\ }\bibfield  {title} {\bibinfo {title} {Quantum velocity limits
  for multiple observables: Conservation laws, correlations, and macroscopic
  systems},\ }\href {https://doi.org/10.1142/S0217979222300079} {\bibfield
  {journal} {\bibinfo  {journal} {arXiv:2305.03190}\ } (\bibinfo {year}
  {2023})}\BibitemShut {NoStop}%
\bibitem [{\citenamefont {Garc\'{\i}a-Pintos}\ \emph
  {et~al.}(2022)\citenamefont {Garc\'{\i}a-Pintos}, \citenamefont {Nicholson},
  \citenamefont {Green}, \citenamefont {del Campo},\ and\ \citenamefont
  {Gorshkov}}]{Pintos2022}%
  \BibitemOpen
  \bibfield  {author} {\bibinfo {author} {\bibfnamefont {L.~P.}\ \bibnamefont
  {Garc\'{\i}a-Pintos}}, \bibinfo {author} {\bibfnamefont {S.~B.}\ \bibnamefont
  {Nicholson}}, \bibinfo {author} {\bibfnamefont {J.~R.}\ \bibnamefont
  {Green}}, \bibinfo {author} {\bibfnamefont {A.}~\bibnamefont {del Campo}},\
  and\ \bibinfo {author} {\bibfnamefont {A.~V.}\ \bibnamefont {Gorshkov}},\
  }\bibfield  {title} {\bibinfo {title} {Unifying quantum and classical speed
  limits on observables},\ }\href {https://doi.org/10.1103/PhysRevX.12.011038}
  {\bibfield  {journal} {\bibinfo  {journal} {Physical Review X}\ }\textbf
  {\bibinfo {volume} {12}},\ \bibinfo {pages} {011038} (\bibinfo {year}
  {2022})}\BibitemShut {NoStop}%
\bibitem [{\citenamefont {Hamazaki}(2022)}]{Hamazaki2022}%
  \BibitemOpen
  \bibfield  {author} {\bibinfo {author} {\bibfnamefont {R.}~\bibnamefont
  {Hamazaki}},\ }\bibfield  {title} {\bibinfo {title} {Speed limits for
  macroscopic transitions},\ }\href
  {https://doi.org/10.1103/PRXQuantum.3.020319} {\bibfield  {journal} {\bibinfo
   {journal} {PRX Quantum}\ }\textbf {\bibinfo {volume} {3}},\ \bibinfo {pages}
  {020319} (\bibinfo {year} {2022})}\BibitemShut {NoStop}%
\bibitem [{\citenamefont {Hörnedal}\ \emph {et~al.}(2023)\citenamefont
  {Hörnedal}, \citenamefont {Carabba}, \citenamefont {Takahashi},\ and\
  \citenamefont {del Campo}}]{Niklas2023}%
  \BibitemOpen
  \bibfield  {author} {\bibinfo {author} {\bibfnamefont {N.}~\bibnamefont
  {Hörnedal}}, \bibinfo {author} {\bibfnamefont {N.}~\bibnamefont {Carabba}},
  \bibinfo {author} {\bibfnamefont {K.}~\bibnamefont {Takahashi}},\ and\
  \bibinfo {author} {\bibfnamefont {A.}~\bibnamefont {del Campo}},\ }\href
  {https://doi.org/10.48550/ARXIV.2301.04372} {\bibinfo {title} {Geometric
  operator quantum speed limit, wegner hamiltonian flow and operator growth}}
  (\bibinfo {year} {2023})\BibitemShut {NoStop}%
\bibitem [{\citenamefont {Carabba}\ \emph {et~al.}(2022)\citenamefont
  {Carabba}, \citenamefont {Hörnedal},\ and\ \citenamefont {del
  Campo}}]{Carabba_2022}%
  \BibitemOpen
  \bibfield  {author} {\bibinfo {author} {\bibfnamefont {N.}~\bibnamefont
  {Carabba}}, \bibinfo {author} {\bibfnamefont {N.}~\bibnamefont {Hörnedal}},\
  and\ \bibinfo {author} {\bibfnamefont {A.}~\bibnamefont {del Campo}},\
  }\bibfield  {title} {\bibinfo {title} {Quantum speed limits on operator flows
  and correlation functions},\ }\href
  {https://doi.org/10.22331/q-2022-12-22-884} {\bibfield  {journal} {\bibinfo
  {journal} {Quantum}\ }\textbf {\bibinfo {volume} {6}},\ \bibinfo {pages}
  {884} (\bibinfo {year} {2022})}\BibitemShut {NoStop}%
\bibitem [{\citenamefont {Hasegawa}(2023)}]{Hasegawa2023}%
  \BibitemOpen
  \bibfield  {author} {\bibinfo {author} {\bibfnamefont {Y.}~\bibnamefont
  {Hasegawa}},\ }\bibfield  {title} {\bibinfo {title} {Thermodynamic
  correlation inequality},\ }\href {https://doi.org/10.48550/arXiv.2301.03060}
  {\bibfield  {journal} {\bibinfo  {journal} {arXiv:2301.03060}\ } (\bibinfo
  {year} {2023})}\BibitemShut {NoStop}%
\bibitem [{\citenamefont {Jing}\ \emph {et~al.}(2016)\citenamefont {Jing},
  \citenamefont {Wu},\ and\ \citenamefont {del Campo}}]{Jing_2016}%
  \BibitemOpen
  \bibfield  {author} {\bibinfo {author} {\bibfnamefont {J.}~\bibnamefont
  {Jing}}, \bibinfo {author} {\bibfnamefont {L.}~\bibnamefont {Wu}},\ and\
  \bibinfo {author} {\bibfnamefont {A.}~\bibnamefont {del Campo}},\ }\bibfield
  {title} {\bibinfo {title} {Fundamental speed limits to the generation of
  quantumness},\ }\bibfield  {journal} {\bibinfo  {journal} {Scientific
  Reports}\ }\textbf {\bibinfo {volume} {6}},\ \href
  {https://doi.org/10.1038/srep38149} {10.1038/srep38149} (\bibinfo {year}
  {2016})\BibitemShut {NoStop}%
\bibitem [{\citenamefont {Mohan}\ \emph {et~al.}(2022)\citenamefont {Mohan},
  \citenamefont {Das},\ and\ \citenamefont {Pati}}]{Brij2022}%
  \BibitemOpen
  \bibfield  {author} {\bibinfo {author} {\bibfnamefont {B.}~\bibnamefont
  {Mohan}}, \bibinfo {author} {\bibfnamefont {S.}~\bibnamefont {Das}},\ and\
  \bibinfo {author} {\bibfnamefont {A.~K.}\ \bibnamefont {Pati}},\ }\bibfield
  {title} {\bibinfo {title} {Quantum speed limits for information and
  coherence},\ }\href {https://doi.org/10.1088/1367-2630/ac753c} {\bibfield
  {journal} {\bibinfo  {journal} {New Journal of Physics}\ }\textbf {\bibinfo
  {volume} {24}},\ \bibinfo {pages} {065003} (\bibinfo {year}
  {2022})}\BibitemShut {NoStop}%
\bibitem [{\citenamefont {Bera}\ \emph {et~al.}(2013)\citenamefont {Bera},
  \citenamefont {Prabhu}, \citenamefont {Pati}, \citenamefont {De},\ and\
  \citenamefont {Sen}}]{Bera2013}%
  \BibitemOpen
  \bibfield  {author} {\bibinfo {author} {\bibfnamefont {M.~N.}\ \bibnamefont
  {Bera}}, \bibinfo {author} {\bibfnamefont {R.}~\bibnamefont {Prabhu}},
  \bibinfo {author} {\bibfnamefont {A.~K.}\ \bibnamefont {Pati}}, \bibinfo
  {author} {\bibfnamefont {A.~S.}\ \bibnamefont {De}},\ and\ \bibinfo {author}
  {\bibfnamefont {U.}~\bibnamefont {Sen}},\ }\bibfield  {title} {\bibinfo
  {title} {Limit on time-energy uncertainty with multipartite entanglement},\
  }\href {https://doi.org/10.48550/arXiv.1303.0706} {\bibfield  {journal}
  {\bibinfo  {journal} {arXiv:1303.0706}\ } (\bibinfo {year}
  {2013})}\BibitemShut {NoStop}%
\bibitem [{\citenamefont {Rudnicki}(2021)}]{Rudnicki20201}%
  \BibitemOpen
  \bibfield  {author} {\bibinfo {author} {\bibfnamefont {L.}~\bibnamefont
  {Rudnicki}},\ }\bibfield  {title} {\bibinfo {title} {Quantum speed limit and
  geometric measure of entanglement},\ }\href
  {https://doi.org/10.1103/PhysRevA.104.032417} {\bibfield  {journal} {\bibinfo
   {journal} {Physical Review A}\ }\textbf {\bibinfo {volume} {104}},\ \bibinfo
  {pages} {032417} (\bibinfo {year} {2021})}\BibitemShut {NoStop}%
\bibitem [{\citenamefont {Pandey}\ \emph {et~al.}(2023)\citenamefont {Pandey},
  \citenamefont {Shrimali}, \citenamefont {Mohan}, \citenamefont {Das},\ and\
  \citenamefont {Pati}}]{vivek2022}%
  \BibitemOpen
  \bibfield  {author} {\bibinfo {author} {\bibfnamefont {V.}~\bibnamefont
  {Pandey}}, \bibinfo {author} {\bibfnamefont {D.}~\bibnamefont {Shrimali}},
  \bibinfo {author} {\bibfnamefont {B.}~\bibnamefont {Mohan}}, \bibinfo
  {author} {\bibfnamefont {S.}~\bibnamefont {Das}},\ and\ \bibinfo {author}
  {\bibfnamefont {A.~K.}\ \bibnamefont {Pati}},\ }\bibfield  {title} {\bibinfo
  {title} {Speed limits on correlations in bipartite quantum systems},\ }\href
  {https://doi.org/10.1103/PhysRevA.107.052419} {\bibfield  {journal} {\bibinfo
   {journal} {Physical Review A}\ }\textbf {\bibinfo {volume} {107}},\ \bibinfo
  {pages} {052419} (\bibinfo {year} {2023})}\BibitemShut {NoStop}%
\bibitem [{\citenamefont {Shrimali}\ \emph {et~al.}(2022)\citenamefont
  {Shrimali}, \citenamefont {Bhowmick}, \citenamefont {Pandey},\ and\
  \citenamefont {Pati}}]{Divyansh2022}%
  \BibitemOpen
  \bibfield  {author} {\bibinfo {author} {\bibfnamefont {D.}~\bibnamefont
  {Shrimali}}, \bibinfo {author} {\bibfnamefont {S.}~\bibnamefont {Bhowmick}},
  \bibinfo {author} {\bibfnamefont {V.}~\bibnamefont {Pandey}},\ and\ \bibinfo
  {author} {\bibfnamefont {A.~K.}\ \bibnamefont {Pati}},\ }\bibfield  {title}
  {\bibinfo {title} {Capacity of entanglement for a nonlocal hamiltonian},\
  }\href {https://doi.org/10.1103/PhysRevA.106.042419} {\bibfield  {journal}
  {\bibinfo  {journal} {Physical Review A}\ }\textbf {\bibinfo {volume}
  {106}},\ \bibinfo {pages} {042419} (\bibinfo {year} {2022})}\BibitemShut
  {NoStop}%
\bibitem [{\citenamefont {Campaioli}\ \emph {et~al.}(2022)\citenamefont
  {Campaioli}, \citenamefont {Yu}, \citenamefont {Pollock},\ and\ \citenamefont
  {Modi}}]{Campaioli_2022}%
  \BibitemOpen
  \bibfield  {author} {\bibinfo {author} {\bibfnamefont {F.}~\bibnamefont
  {Campaioli}}, \bibinfo {author} {\bibfnamefont {C.}~\bibnamefont {Yu}},
  \bibinfo {author} {\bibfnamefont {F.~A.}\ \bibnamefont {Pollock}},\ and\
  \bibinfo {author} {\bibfnamefont {K.}~\bibnamefont {Modi}},\ }\bibfield
  {title} {\bibinfo {title} {Resource speed limits: maximal rate of resource
  variation},\ }\href {https://doi.org/10.1088/1367-2630/ac7346} {\bibfield
  {journal} {\bibinfo  {journal} {New Journal of Physics}\ }\textbf {\bibinfo
  {volume} {24}},\ \bibinfo {pages} {065001} (\bibinfo {year}
  {2022})}\BibitemShut {NoStop}%
\bibitem [{\citenamefont {Pires}(2022)}]{Diego2022}%
  \BibitemOpen
  \bibfield  {author} {\bibinfo {author} {\bibfnamefont {D.~P.}\ \bibnamefont
  {Pires}},\ }\bibfield  {title} {\bibinfo {title} {Unified entropies and
  quantum speed limits for nonunitary dynamics},\ }\href
  {https://doi.org/10.1103/PhysRevA.106.012403} {\bibfield  {journal} {\bibinfo
   {journal} {Physical Review A}\ }\textbf {\bibinfo {volume} {106}},\ \bibinfo
  {pages} {012403} (\bibinfo {year} {2022})}\BibitemShut {NoStop}%
\bibitem [{\citenamefont {Pires}\ \emph {et~al.}(2021)\citenamefont {Pires},
  \citenamefont {Modi},\ and\ \citenamefont {C{\'{e} }leri}}]{Pires2021}%
  \BibitemOpen
  \bibfield  {author} {\bibinfo {author} {\bibfnamefont {D.~P.}\ \bibnamefont
  {Pires}}, \bibinfo {author} {\bibfnamefont {K.}~\bibnamefont {Modi}},\ and\
  \bibinfo {author} {\bibfnamefont {L.~C.}\ \bibnamefont {C{\'{e} }leri}},\
  }\bibfield  {title} {\bibinfo {title} {Bounding generalized relative
  entropies: Nonasymptotic quantum speed limits},\ }\bibfield  {journal}
  {\bibinfo  {journal} {Physical Review E}\ }\textbf {\bibinfo {volume}
  {103}},\ \href {https://doi.org/10.1103/physreve.103.032105}
  {10.1103/physreve.103.032105} (\bibinfo {year} {2021})\BibitemShut {NoStop}%
\bibitem [{\citenamefont {Paulson}\ and\ \citenamefont
  {Banerjee}(2022)}]{Paulson2022}%
  \BibitemOpen
  \bibfield  {author} {\bibinfo {author} {\bibfnamefont {K.~G.}\ \bibnamefont
  {Paulson}}\ and\ \bibinfo {author} {\bibfnamefont {S.}~\bibnamefont
  {Banerjee}},\ }\href {https://doi.org/10.48550/ARXIV.2205.11882} {\bibinfo
  {title} {Quantum speed limit for the creation and decay of quantum
  correlations}} (\bibinfo {year} {2022})\BibitemShut {NoStop}%
\bibitem [{\citenamefont {Shaham}\ \emph {et~al.}(2015)\citenamefont {Shaham},
  \citenamefont {Halevy}, \citenamefont {Dovrat}, \citenamefont {Megidish},\
  and\ \citenamefont {Eisenberg}}]{Shaham_2015}%
  \BibitemOpen
  \bibfield  {author} {\bibinfo {author} {\bibfnamefont {A.}~\bibnamefont
  {Shaham}}, \bibinfo {author} {\bibfnamefont {A.}~\bibnamefont {Halevy}},
  \bibinfo {author} {\bibfnamefont {L.}~\bibnamefont {Dovrat}}, \bibinfo
  {author} {\bibfnamefont {E.}~\bibnamefont {Megidish}},\ and\ \bibinfo
  {author} {\bibfnamefont {H.~S.}\ \bibnamefont {Eisenberg}},\ }\bibfield
  {title} {\bibinfo {title} {Entanglement dynamics in the presence of
  controlled unital noise},\ }\href
  {https://doi.org/https://doi.org/10.1038/srep10796} {\bibfield  {journal}
  {\bibinfo  {journal} {Scientific Reports}\ }\textbf {\bibinfo {volume} {5}},\
  \bibinfo {pages} {10796} (\bibinfo {year} {2015})}\BibitemShut {NoStop}%
\bibitem [{\citenamefont {Nosrati}\ \emph {et~al.}(2020)\citenamefont
  {Nosrati}, \citenamefont {Castellini}, \citenamefont {Compagno},\ and\
  \citenamefont {Franco}}]{Nosrati_2020}%
  \BibitemOpen
  \bibfield  {author} {\bibinfo {author} {\bibfnamefont {F.}~\bibnamefont
  {Nosrati}}, \bibinfo {author} {\bibfnamefont {A.}~\bibnamefont {Castellini}},
  \bibinfo {author} {\bibfnamefont {G.}~\bibnamefont {Compagno}},\ and\
  \bibinfo {author} {\bibfnamefont {R.~L.}\ \bibnamefont {Franco}},\ }\bibfield
   {title} {\bibinfo {title} {Robust entanglement preparation against noise by
  controlling spatial indistinguishability},\ }\bibfield  {journal} {\bibinfo
  {journal} {npj Quantum Information}\ }\textbf {\bibinfo {volume} {6}},\ \href
  {https://doi.org/10.1038/s41534-020-0271-7} {10.1038/s41534-020-0271-7}
  (\bibinfo {year} {2020})\BibitemShut {NoStop}%
\bibitem [{\citenamefont {Schachenmayer}\ \emph {et~al.}(2013)\citenamefont
  {Schachenmayer}, \citenamefont {Lanyon}, \citenamefont {Roos},\ and\
  \citenamefont {Daley}}]{Laynon2013}%
  \BibitemOpen
  \bibfield  {author} {\bibinfo {author} {\bibfnamefont {J.}~\bibnamefont
  {Schachenmayer}}, \bibinfo {author} {\bibfnamefont {B.~P.}\ \bibnamefont
  {Lanyon}}, \bibinfo {author} {\bibfnamefont {C.~F.}\ \bibnamefont {Roos}},\
  and\ \bibinfo {author} {\bibfnamefont {A.~J.}\ \bibnamefont {Daley}},\
  }\bibfield  {title} {\bibinfo {title} {Entanglement growth in quench dynamics
  with variable range interactions},\ }\href
  {https://doi.org/10.1103/PhysRevX.3.031015} {\bibfield  {journal} {\bibinfo
  {journal} {Physical Review X}\ }\textbf {\bibinfo {volume} {3}},\ \bibinfo
  {pages} {031015} (\bibinfo {year} {2013})}\BibitemShut {NoStop}%
\bibitem [{\citenamefont {Hamilton}\ and\ \citenamefont
  {Clark}(2023)}]{Hamilton2023}%
  \BibitemOpen
  \bibfield  {author} {\bibinfo {author} {\bibfnamefont {G.~A.}\ \bibnamefont
  {Hamilton}}\ and\ \bibinfo {author} {\bibfnamefont {B.~K.}\ \bibnamefont
  {Clark}},\ }\bibfield  {title} {\bibinfo {title} {Quantifying unitary flow
  efficiency and entanglement for many-body localization},\ }\href
  {https://doi.org/10.1103/PhysRevB.107.064203} {\bibfield  {journal} {\bibinfo
   {journal} {Physical Review B}\ }\textbf {\bibinfo {volume} {107}},\ \bibinfo
  {pages} {064203} (\bibinfo {year} {2023})}\BibitemShut {NoStop}%
\bibitem [{\citenamefont {Horodecki}\ \emph {et~al.}(1998)\citenamefont
  {Horodecki}, \citenamefont {Horodecki},\ and\ \citenamefont
  {Horodecki}}]{Horodicki1998}%
  \BibitemOpen
  \bibfield  {author} {\bibinfo {author} {\bibfnamefont {M.}~\bibnamefont
  {Horodecki}}, \bibinfo {author} {\bibfnamefont {P.}~\bibnamefont
  {Horodecki}},\ and\ \bibinfo {author} {\bibfnamefont {R.}~\bibnamefont
  {Horodecki}},\ }\bibfield  {title} {\bibinfo {title} {Mixed-state
  entanglement and distillation: Is there a ``bound'' entanglement in
  nature?},\ }\href {https://doi.org/10.1103/PhysRevLett.80.5239} {\bibfield
  {journal} {\bibinfo  {journal} {Physical Review Letters}\ }\textbf {\bibinfo
  {volume} {80}},\ \bibinfo {pages} {5239} (\bibinfo {year}
  {1998})}\BibitemShut {NoStop}%
\bibitem [{\citenamefont {Vedral}\ \emph {et~al.}(1997)\citenamefont {Vedral},
  \citenamefont {Plenio}, \citenamefont {Rippin},\ and\ \citenamefont
  {Knight}}]{Vedral1997}%
  \BibitemOpen
  \bibfield  {author} {\bibinfo {author} {\bibfnamefont {V.}~\bibnamefont
  {Vedral}}, \bibinfo {author} {\bibfnamefont {M.~B.}\ \bibnamefont {Plenio}},
  \bibinfo {author} {\bibfnamefont {M.~A.}\ \bibnamefont {Rippin}},\ and\
  \bibinfo {author} {\bibfnamefont {P.~L.}\ \bibnamefont {Knight}},\ }\bibfield
   {title} {\bibinfo {title} {Quantifying entanglement},\ }\href
  {https://doi.org/10.1103/PhysRevLett.78.2275} {\bibfield  {journal} {\bibinfo
   {journal} {Physical Review Letters}\ }\textbf {\bibinfo {volume} {78}},\
  \bibinfo {pages} {2275} (\bibinfo {year} {1997})}\BibitemShut {NoStop}%
\bibitem [{\citenamefont {Eisert}\ \emph {et~al.}(2003)\citenamefont {Eisert},
  \citenamefont {Audenaert},\ and\ \citenamefont {Plenio}}]{Eisert_2003}%
  \BibitemOpen
  \bibfield  {author} {\bibinfo {author} {\bibfnamefont {J.}~\bibnamefont
  {Eisert}}, \bibinfo {author} {\bibfnamefont {K.}~\bibnamefont {Audenaert}},\
  and\ \bibinfo {author} {\bibfnamefont {M.~B.}\ \bibnamefont {Plenio}},\
  }\bibfield  {title} {\bibinfo {title} {Remarks on entanglement measures and
  non-local state distinguishability},\ }\href
  {https://doi.org/10.1088/0305-4470/36/20/316} {\bibfield  {journal} {\bibinfo
   {journal} {Journal of Physics A: Mathematical and General}\ }\textbf
  {\bibinfo {volume} {36}},\ \bibinfo {pages} {5605} (\bibinfo {year}
  {2003})}\BibitemShut {NoStop}%
\bibitem [{\citenamefont {Werner}(1989)}]{Werner1989}%
  \BibitemOpen
  \bibfield  {author} {\bibinfo {author} {\bibfnamefont {R.~F.}\ \bibnamefont
  {Werner}},\ }\bibfield  {title} {\bibinfo {title} {Quantum states with
  einstein-podolsky-rosen correlations admitting a hidden-variable model},\
  }\href {https://doi.org/10.1103/PhysRevA.40.4277} {\bibfield  {journal}
  {\bibinfo  {journal} {Physical Review A}\ }\textbf {\bibinfo {volume} {40}},\
  \bibinfo {pages} {4277} (\bibinfo {year} {1989})}\BibitemShut {NoStop}%
\bibitem [{\citenamefont {Bennett}\ \emph {et~al.}(1996)\citenamefont
  {Bennett}, \citenamefont {DiVincenzo}, \citenamefont {Smolin},\ and\
  \citenamefont {Wootters}}]{Bennett1996}%
  \BibitemOpen
  \bibfield  {author} {\bibinfo {author} {\bibfnamefont {C.~H.}\ \bibnamefont
  {Bennett}}, \bibinfo {author} {\bibfnamefont {D.~P.}\ \bibnamefont
  {DiVincenzo}}, \bibinfo {author} {\bibfnamefont {J.~A.}\ \bibnamefont
  {Smolin}},\ and\ \bibinfo {author} {\bibfnamefont {W.~K.}\ \bibnamefont
  {Wootters}},\ }\bibfield  {title} {\bibinfo {title} {Mixed-state entanglement
  and quantum error correction},\ }\href
  {https://doi.org/10.1103/PhysRevA.54.3824} {\bibfield  {journal} {\bibinfo
  {journal} {Physical Review A}\ }\textbf {\bibinfo {volume} {54}},\ \bibinfo
  {pages} {3824} (\bibinfo {year} {1996})}\BibitemShut {NoStop}%
\bibitem [{\citenamefont {Wootters}(2001)}]{Wootters2001}%
  \BibitemOpen
  \bibfield  {author} {\bibinfo {author} {\bibfnamefont {W.~K.}\ \bibnamefont
  {Wootters}},\ }\bibfield  {title} {\bibinfo {title} {Entanglement of
  formation and concurrence},\ }\href
  {https://dl.acm.org/doi/10.5555/2011326.2011329} {\bibfield  {journal}
  {\bibinfo  {journal} {Quantum Information and Computation}\ }\textbf
  {\bibinfo {volume} {1}},\ \bibinfo {pages} {27} (\bibinfo {year}
  {2001})}\BibitemShut {NoStop}%
\bibitem [{\citenamefont {Wootters}(1998)}]{Wootters1998}%
  \BibitemOpen
  \bibfield  {author} {\bibinfo {author} {\bibfnamefont {W.~K.}\ \bibnamefont
  {Wootters}},\ }\bibfield  {title} {\bibinfo {title} {Entanglement of
  formation of an arbitrary state of two qubits},\ }\href
  {https://doi.org/10.1103/PhysRevLett.80.2245} {\bibfield  {journal} {\bibinfo
   {journal} {Physical Review Letters}\ }\textbf {\bibinfo {volume} {80}},\
  \bibinfo {pages} {2245} (\bibinfo {year} {1998})}\BibitemShut {NoStop}%
\bibitem [{\citenamefont {Vidal}\ and\ \citenamefont
  {Werner}(2002)}]{Vidal2002}%
  \BibitemOpen
  \bibfield  {author} {\bibinfo {author} {\bibfnamefont {G.}~\bibnamefont
  {Vidal}}\ and\ \bibinfo {author} {\bibfnamefont {R.~F.}\ \bibnamefont
  {Werner}},\ }\bibfield  {title} {\bibinfo {title} {Computable measure of
  entanglement},\ }\href {https://doi.org/10.1103/PhysRevA.65.032314}
  {\bibfield  {journal} {\bibinfo  {journal} {Physical Review A}\ }\textbf
  {\bibinfo {volume} {65}},\ \bibinfo {pages} {032314} (\bibinfo {year}
  {2002})}\BibitemShut {NoStop}%
\bibitem [{\citenamefont {Vedral}\ and\ \citenamefont {Plenio}(1998)}]{Vedral}%
  \BibitemOpen
  \bibfield  {author} {\bibinfo {author} {\bibfnamefont {V.}~\bibnamefont
  {Vedral}}\ and\ \bibinfo {author} {\bibfnamefont {M.~B.}\ \bibnamefont
  {Plenio}},\ }\bibfield  {title} {\bibinfo {title} {Entanglement measures and
  purification procedures},\ }\href {https://doi.org/10.1103/PhysRevA.57.1619}
  {\bibfield  {journal} {\bibinfo  {journal} {Physical Review A}\ }\textbf
  {\bibinfo {volume} {57}},\ \bibinfo {pages} {1619} (\bibinfo {year}
  {1998})}\BibitemShut {NoStop}%
\bibitem [{\citenamefont {Friedland}\ and\ \citenamefont
  {Gour}(2011)}]{Fried2011}%
  \BibitemOpen
  \bibfield  {author} {\bibinfo {author} {\bibfnamefont {S.}~\bibnamefont
  {Friedland}}\ and\ \bibinfo {author} {\bibfnamefont {G.}~\bibnamefont
  {Gour}},\ }\bibfield  {title} {\bibinfo {title} {An explicit expression for
  the relative entropy of entanglement in all dimensions},\ }\href
  {https://doi.org/10.1063/1.3591132} {\bibfield  {journal} {\bibinfo
  {journal} {Journal of Mathematical Physics}\ }\textbf {\bibinfo {volume}
  {52}},\ \bibinfo {pages} {052201} (\bibinfo {year} {2011})}\BibitemShut
  {NoStop}%
\bibitem [{\citenamefont {Das}\ \emph {et~al.}(2021)\citenamefont {Das},
  \citenamefont {B\"auml}, \citenamefont {Winczewski},\ and\ \citenamefont
  {Horodecki}}]{Das2021}%
  \BibitemOpen
  \bibfield  {author} {\bibinfo {author} {\bibfnamefont {S.}~\bibnamefont
  {Das}}, \bibinfo {author} {\bibfnamefont {S.}~\bibnamefont {B\"auml}},
  \bibinfo {author} {\bibfnamefont {M.}~\bibnamefont {Winczewski}},\ and\
  \bibinfo {author} {\bibfnamefont {K.}~\bibnamefont {Horodecki}},\ }\bibfield
  {title} {\bibinfo {title} {Universal limitations on quantum key distribution
  over a network},\ }\href {https://doi.org/10.1103/PhysRevX.11.041016}
  {\bibfield  {journal} {\bibinfo  {journal} {Physical Review X}\ }\textbf
  {\bibinfo {volume} {11}},\ \bibinfo {pages} {041016} (\bibinfo {year}
  {2021})}\BibitemShut {NoStop}%
\bibitem [{\citenamefont {Yu}\ and\ \citenamefont {Eberly}(2006)}]{Yu2006}%
  \BibitemOpen
  \bibfield  {author} {\bibinfo {author} {\bibfnamefont {T.}~\bibnamefont
  {Yu}}\ and\ \bibinfo {author} {\bibfnamefont {J.~H.}\ \bibnamefont
  {Eberly}},\ }\bibfield  {title} {\bibinfo {title} {Quantum open system
  theory: Bipartite aspects},\ }\href
  {https://doi.org/10.1103/PhysRevLett.97.140403} {\bibfield  {journal}
  {\bibinfo  {journal} {Phys. Rev. Lett.}\ }\textbf {\bibinfo {volume} {97}},\
  \bibinfo {pages} {140403} (\bibinfo {year} {2006})}\BibitemShut {NoStop}%
\bibitem [{\citenamefont {Gatto}\ \emph {et~al.}(2019)\citenamefont {Gatto},
  \citenamefont {De~Pasquale},\ and\ \citenamefont {Giovannetti}}]{Gatto2019}%
  \BibitemOpen
  \bibfield  {author} {\bibinfo {author} {\bibfnamefont {D.}~\bibnamefont
  {Gatto}}, \bibinfo {author} {\bibfnamefont {A.}~\bibnamefont {De~Pasquale}},\
  and\ \bibinfo {author} {\bibfnamefont {V.}~\bibnamefont {Giovannetti}},\
  }\bibfield  {title} {\bibinfo {title} {Degradation of entanglement in
  markovian noise},\ }\href {https://doi.org/10.1103/PhysRevA.99.032307}
  {\bibfield  {journal} {\bibinfo  {journal} {Phys. Rev. A}\ }\textbf {\bibinfo
  {volume} {99}},\ \bibinfo {pages} {032307} (\bibinfo {year}
  {2019})}\BibitemShut {NoStop}%
\bibitem [{\citenamefont {Sakuldee}\ and\ \citenamefont
  {Rudnicki}(2023{\natexlab{a}})}]{Sakuldee2023}%
  \BibitemOpen
  \bibfield  {author} {\bibinfo {author} {\bibfnamefont {F.}~\bibnamefont
  {Sakuldee}}\ and\ \bibinfo {author} {\bibfnamefont {L.}~\bibnamefont
  {Rudnicki}},\ }\bibfield  {title} {\bibinfo {title} {Bounds on the breaking
  time for entanglement-breaking channels},\ }\href
  {https://doi.org/10.1103/PhysRevA.107.022430} {\bibfield  {journal} {\bibinfo
   {journal} {Phys. Rev. A}\ }\textbf {\bibinfo {volume} {107}},\ \bibinfo
  {pages} {022430} (\bibinfo {year} {2023}{\natexlab{a}})}\BibitemShut
  {NoStop}%
\bibitem [{\citenamefont {Sorelli}\ \emph {et~al.}(2019)\citenamefont
  {Sorelli}, \citenamefont {Gessner}, \citenamefont {Smerzi},\ and\
  \citenamefont {Pezz\`e}}]{Sorelli2019}%
  \BibitemOpen
  \bibfield  {author} {\bibinfo {author} {\bibfnamefont {G.}~\bibnamefont
  {Sorelli}}, \bibinfo {author} {\bibfnamefont {M.}~\bibnamefont {Gessner}},
  \bibinfo {author} {\bibfnamefont {A.}~\bibnamefont {Smerzi}},\ and\ \bibinfo
  {author} {\bibfnamefont {L.}~\bibnamefont {Pezz\`e}},\ }\bibfield  {title}
  {\bibinfo {title} {Fast and optimal generation of entanglement in bosonic
  josephson junctions},\ }\href {https://doi.org/10.1103/PhysRevA.99.022329}
  {\bibfield  {journal} {\bibinfo  {journal} {Phys. Rev. A}\ }\textbf {\bibinfo
  {volume} {99}},\ \bibinfo {pages} {022329} (\bibinfo {year}
  {2019})}\BibitemShut {NoStop}%
\bibitem [{\citenamefont {Cieśliński}\ \emph {et~al.}(2023)\citenamefont
  {Cieśliński}, \citenamefont {Kłobus}, \citenamefont {Kurzyński},
  \citenamefont {Paterek},\ and\ \citenamefont {Laskowski}}]{PaweL2023}%
  \BibitemOpen
  \bibfield  {author} {\bibinfo {author} {\bibfnamefont {P.}~\bibnamefont
  {Cieśliński}}, \bibinfo {author} {\bibfnamefont {W.}~\bibnamefont
  {Kłobus}}, \bibinfo {author} {\bibfnamefont {P.}~\bibnamefont {Kurzyński}},
  \bibinfo {author} {\bibfnamefont {T.}~\bibnamefont {Paterek}},\ and\ \bibinfo
  {author} {\bibfnamefont {W.}~\bibnamefont {Laskowski}},\ }\bibfield  {title}
  {\bibinfo {title} {The fastest generation of multipartite entanglement with
  natural interactions},\ }\href {https://doi.org/10.1088/1367-2630/acf953}
  {\bibfield  {journal} {\bibinfo  {journal} {New Journal of Physics}\ }\textbf
  {\bibinfo {volume} {25}},\ \bibinfo {pages} {093040} (\bibinfo {year}
  {2023})}\BibitemShut {NoStop}%
\bibitem [{\citenamefont {Bravyi}(2007)}]{Bravyi2007}%
  \BibitemOpen
  \bibfield  {author} {\bibinfo {author} {\bibfnamefont {S.}~\bibnamefont
  {Bravyi}},\ }\bibfield  {title} {\bibinfo {title} {Upper bounds on entangling
  rates of bipartite hamiltonians},\ }\href
  {https://doi.org/10.1103/PhysRevA.76.052319} {\bibfield  {journal} {\bibinfo
  {journal} {Phys. Rev. A}\ }\textbf {\bibinfo {volume} {76}},\ \bibinfo
  {pages} {052319} (\bibinfo {year} {2007})}\BibitemShut {NoStop}%
\bibitem [{\citenamefont {Gong}\ and\ \citenamefont
  {Hamazaki}(2022{\natexlab{b}})}]{Gong2022}%
  \BibitemOpen
  \bibfield  {author} {\bibinfo {author} {\bibfnamefont {Z.}~\bibnamefont
  {Gong}}\ and\ \bibinfo {author} {\bibfnamefont {R.}~\bibnamefont
  {Hamazaki}},\ }\bibfield  {title} {\bibinfo {title} {Bounds in nonequilibrium
  quantum dynamics},\ }\href {https://doi.org/10.1142/S0217979222300079}
  {\bibfield  {journal} {\bibinfo  {journal} {International Journal of Modern
  Physics B}\ }\textbf {\bibinfo {volume} {36}},\ \bibinfo {pages} {2230007}
  (\bibinfo {year} {2022}{\natexlab{b}})},\ \Eprint
  {https://arxiv.org/abs/https://doi.org/10.1142/S0217979222300079}
  {https://doi.org/10.1142/S0217979222300079} \BibitemShut {NoStop}%
\bibitem [{\citenamefont {Das}\ \emph {et~al.}(2018)\citenamefont {Das},
  \citenamefont {Khatri}, \citenamefont {Siopsis},\ and\ \citenamefont
  {Wilde}}]{Das2018}%
  \BibitemOpen
  \bibfield  {author} {\bibinfo {author} {\bibfnamefont {S.}~\bibnamefont
  {Das}}, \bibinfo {author} {\bibfnamefont {S.}~\bibnamefont {Khatri}},
  \bibinfo {author} {\bibfnamefont {G.}~\bibnamefont {Siopsis}},\ and\ \bibinfo
  {author} {\bibfnamefont {M.~M.}\ \bibnamefont {Wilde}},\ }\bibfield  {title}
  {\bibinfo {title} {Fundamental limits on quantum dynamics based on entropy
  change},\ }\href {https://doi.org/10.1063/1.4997044} {\bibfield  {journal}
  {\bibinfo  {journal} {Journal of Mathematical Physics}\ }\textbf {\bibinfo
  {volume} {59}},\ \bibinfo {pages} {012205} (\bibinfo {year}
  {2018})}\BibitemShut {NoStop}%
\bibitem [{\citenamefont {Campaioli}\ \emph {et~al.}(2019)\citenamefont
  {Campaioli}, \citenamefont {Pollock},\ and\ \citenamefont
  {Modi}}]{Campaioli_2019}%
  \BibitemOpen
  \bibfield  {author} {\bibinfo {author} {\bibfnamefont {F.}~\bibnamefont
  {Campaioli}}, \bibinfo {author} {\bibfnamefont {F.~A.}\ \bibnamefont
  {Pollock}},\ and\ \bibinfo {author} {\bibfnamefont {K.}~\bibnamefont
  {Modi}},\ }\bibfield  {title} {\bibinfo {title} {Tight, robust, and feasible
  quantum speed limits for open dynamics},\ }\href
  {https://doi.org/10.22331/q-2019-08-05-168} {\bibfield  {journal} {\bibinfo
  {journal} {Quantum}\ }\textbf {\bibinfo {volume} {3}},\ \bibinfo {pages}
  {168} (\bibinfo {year} {2019})}\BibitemShut {NoStop}%
\bibitem [{\citenamefont {Sakuldee}\ and\ \citenamefont
  {Rudnicki}(2023{\natexlab{b}})}]{Lukasz2023}%
  \BibitemOpen
  \bibfield  {author} {\bibinfo {author} {\bibfnamefont {F.}~\bibnamefont
  {Sakuldee}}\ and\ \bibinfo {author} {\bibfnamefont {L.}~\bibnamefont
  {Rudnicki}},\ }\bibfield  {title} {\bibinfo {title} {Bounds on the breaking
  time for entanglement-breaking channels},\ }\href
  {https://doi.org/10.1103/PhysRevA.107.022430} {\bibfield  {journal} {\bibinfo
   {journal} {Physical Review A}\ }\textbf {\bibinfo {volume} {107}},\ \bibinfo
  {pages} {022430} (\bibinfo {year} {2023}{\natexlab{b}})}\BibitemShut
  {NoStop}%
\bibitem [{\citenamefont {Lindblad}(1976)}]{Lindblad1976}%
  \BibitemOpen
  \bibfield  {author} {\bibinfo {author} {\bibfnamefont {G.}~\bibnamefont
  {Lindblad}},\ }\bibfield  {title} {\bibinfo {title} {On the generators of
  quantum dynamical semigroups},\ }\href {https://doi.org/10.1007/BF01608499}
  {\bibfield  {journal} {\bibinfo  {journal} {Communications in Mathematical
  Physics}\ }\textbf {\bibinfo {volume} {48}},\ \bibinfo {pages} {119}
  (\bibinfo {year} {1976})}\BibitemShut {NoStop}%
\bibitem [{\citenamefont {Gorini}\ \emph {et~al.}(1976)\citenamefont {Gorini},
  \citenamefont {Kossakowski}, \citenamefont {Sudarshan},\ and\ \citenamefont
  {Chandy}}]{Gorini1975}%
  \BibitemOpen
  \bibfield  {author} {\bibinfo {author} {\bibfnamefont {V.}~\bibnamefont
  {Gorini}}, \bibinfo {author} {\bibfnamefont {A.}~\bibnamefont {Kossakowski}},
  \bibinfo {author} {\bibfnamefont {G.}~\bibnamefont {Sudarshan}},\ and\
  \bibinfo {author} {\bibfnamefont {E.}~\bibnamefont {Chandy}},\ }\bibfield
  {title} {\bibinfo {title} {Completely positive dynamical semigroups of
  {N}‐level systems},\ }\href {https://doi.org/10.1063/1.522979} {\bibfield
  {journal} {\bibinfo  {journal} {Journal of Mathematical Physics}\ }\textbf
  {\bibinfo {volume} {17}},\ \bibinfo {pages} {821} (\bibinfo {year}
  {1976})}\BibitemShut {NoStop}%
\bibitem [{\citenamefont {D\"ur}\ \emph {et~al.}(2001)\citenamefont {D\"ur},
  \citenamefont {Vidal}, \citenamefont {Cirac}, \citenamefont {Linden},\ and\
  \citenamefont {Popescu}}]{Vidal2001}%
  \BibitemOpen
  \bibfield  {author} {\bibinfo {author} {\bibfnamefont {W.}~\bibnamefont
  {D\"ur}}, \bibinfo {author} {\bibfnamefont {G.}~\bibnamefont {Vidal}},
  \bibinfo {author} {\bibfnamefont {J.~I.}\ \bibnamefont {Cirac}}, \bibinfo
  {author} {\bibfnamefont {N.}~\bibnamefont {Linden}},\ and\ \bibinfo {author}
  {\bibfnamefont {S.}~\bibnamefont {Popescu}},\ }\bibfield  {title} {\bibinfo
  {title} {Entanglement capabilities of nonlocal hamiltonians},\ }\href
  {https://doi.org/10.1103/PhysRevLett.87.137901} {\bibfield  {journal}
  {\bibinfo  {journal} {Physical Review Letters}\ }\textbf {\bibinfo {volume}
  {87}},\ \bibinfo {pages} {137901} (\bibinfo {year} {2001})}\BibitemShut
  {NoStop}%
\bibitem [{\citenamefont {Mari{\"e}n}\ \emph {et~al.}(2016)\citenamefont
  {Mari{\"e}n}, \citenamefont {Audenaert}, \citenamefont {Van~Acoleyen},\ and\
  \citenamefont {Verstraete}}]{marien2016}%
  \BibitemOpen
  \bibfield  {author} {\bibinfo {author} {\bibfnamefont {M.}~\bibnamefont
  {Mari{\"e}n}}, \bibinfo {author} {\bibfnamefont {K.~M.}\ \bibnamefont
  {Audenaert}}, \bibinfo {author} {\bibfnamefont {K.}~\bibnamefont
  {Van~Acoleyen}},\ and\ \bibinfo {author} {\bibfnamefont {F.}~\bibnamefont
  {Verstraete}},\ }\bibfield  {title} {\bibinfo {title} {Entanglement rates and
  the stability of the area law for the entanglement entropy},\ }\href
  {https://doi.org/https://doi.org/10.1007/s00220-016-2709-5} {\bibfield
  {journal} {\bibinfo  {journal} {Communications in Mathematical Physics}\
  }\textbf {\bibinfo {volume} {346}},\ \bibinfo {pages} {35} (\bibinfo {year}
  {2016})}\BibitemShut {NoStop}%
\bibitem [{\citenamefont {Vershynina}(2019)}]{Vershynina_2019}%
  \BibitemOpen
  \bibfield  {author} {\bibinfo {author} {\bibfnamefont {A.}~\bibnamefont
  {Vershynina}},\ }\bibfield  {title} {\bibinfo {title} {Entanglement rates for
  r{\'{e}}nyi, tsallis, and other entropies},\ }\href
  {https://doi.org/10.1063/1.5037802} {\bibfield  {journal} {\bibinfo
  {journal} {Journal of Mathematical Physics}\ }\textbf {\bibinfo {volume}
  {60}},\ \bibinfo {pages} {022201} (\bibinfo {year} {2019})}\BibitemShut
  {NoStop}%
\bibitem [{\citenamefont {Vershynina}(2015)}]{Anna2015}%
  \BibitemOpen
  \bibfield  {author} {\bibinfo {author} {\bibfnamefont {A.}~\bibnamefont
  {Vershynina}},\ }\bibfield  {title} {\bibinfo {title} {Entanglement rates for
  bipartite open systems},\ }\href {https://doi.org/10.1103/PhysRevA.92.022311}
  {\bibfield  {journal} {\bibinfo  {journal} {Physical Review A}\ }\textbf
  {\bibinfo {volume} {92}},\ \bibinfo {pages} {022311} (\bibinfo {year}
  {2015})}\BibitemShut {NoStop}%
\bibitem [{\citenamefont {Rivas}\ and\ \citenamefont
  {Huelga}(2012)}]{Rivas2012}%
  \BibitemOpen
  \bibfield  {author} {\bibinfo {author} {\bibfnamefont {A.}~\bibnamefont
  {Rivas}}\ and\ \bibinfo {author} {\bibfnamefont {S.~F.}\ \bibnamefont
  {Huelga}},\ }\href@noop {} {\emph {\bibinfo {title} {Open quantum
  systems}}},\ Vol.~\bibinfo {volume} {10}\ (\bibinfo  {publisher} {Springer},\
  \bibinfo {year} {2012})\BibitemShut {NoStop}%
\bibitem [{\citenamefont {Umegaki}(1962)}]{Umegaki1962}%
  \BibitemOpen
  \bibfield  {author} {\bibinfo {author} {\bibfnamefont {H.}~\bibnamefont
  {Umegaki}},\ }\bibfield  {title} {\bibinfo {title} {{Conditional expectation
  in an operator algebra. IV. Entropy and information}},\ }\href
  {https://doi.org/10.2996/kmj/1138844604} {\bibfield  {journal} {\bibinfo
  {journal} {Kodai Mathematical Seminar Reports}\ }\textbf {\bibinfo {volume}
  {14}},\ \bibinfo {pages} {59 } (\bibinfo {year} {1962})}\BibitemShut
  {NoStop}%
\bibitem [{\citenamefont {Luo}(2005)}]{Luo2005}%
  \BibitemOpen
  \bibfield  {author} {\bibinfo {author} {\bibfnamefont {S.}~\bibnamefont
  {Luo}},\ }\bibfield  {title} {\bibinfo {title} {Heisenberg uncertainty
  relation for mixed states},\ }\href
  {https://doi.org/10.1103/PhysRevA.72.042110} {\bibfield  {journal} {\bibinfo
  {journal} {Physical Review A}\ }\textbf {\bibinfo {volume} {72}},\ \bibinfo
  {pages} {042110} (\bibinfo {year} {2005})}\BibitemShut {NoStop}%
\bibitem [{\citenamefont {Robertson}(1929)}]{robertson_uncertain}%
  \BibitemOpen
  \bibfield  {author} {\bibinfo {author} {\bibfnamefont {H.~P.}\ \bibnamefont
  {Robertson}},\ }\bibfield  {title} {\bibinfo {title} {The uncertainty
  principle},\ }\href {https://doi.org/10.1103/PhysRev.34.163} {\bibfield
  {journal} {\bibinfo  {journal} {Phys. Rev.}\ }\textbf {\bibinfo {volume}
  {34}},\ \bibinfo {pages} {163} (\bibinfo {year} {1929})}\BibitemShut
  {NoStop}%
\bibitem [{\citenamefont {Cirac}\ \emph {et~al.}(2001)\citenamefont {Cirac},
  \citenamefont {Dür}, \citenamefont {Kraus},\ and\ \citenamefont
  {Lewenstein}}]{Cirac_2001}%
  \BibitemOpen
  \bibfield  {author} {\bibinfo {author} {\bibfnamefont {J.~I.}\ \bibnamefont
  {Cirac}}, \bibinfo {author} {\bibfnamefont {W.}~\bibnamefont {Dür}},
  \bibinfo {author} {\bibfnamefont {B.}~\bibnamefont {Kraus}},\ and\ \bibinfo
  {author} {\bibfnamefont {M.}~\bibnamefont {Lewenstein}},\ }\bibfield  {title}
  {\bibinfo {title} {Entangling operations and their implementation using a
  small amount of entanglement},\ }\href
  {https://doi.org/10.1103/physrevlett.86.544} {\bibfield  {journal} {\bibinfo
  {journal} {Physical Review Letters}\ }\textbf {\bibinfo {volume} {86}},\
  \bibinfo {pages} {544} (\bibinfo {year} {2001})}\BibitemShut {NoStop}%
\bibitem [{\citenamefont {Kato}(2013)}]{kato2013perturbation}%
  \BibitemOpen
  \bibfield  {author} {\bibinfo {author} {\bibfnamefont {T.}~\bibnamefont
  {Kato}},\ }\href@noop {} {\emph {\bibinfo {title} {Perturbation theory for
  linear operators}}},\ Vol.\ \bibinfo {volume} {132}\ (\bibinfo  {publisher}
  {Springer Science \& Business Media},\ \bibinfo {year} {2013})\BibitemShut
  {NoStop}%
\bibitem [{\citenamefont {Deffner}(2017)}]{Deffner2017}%
  \BibitemOpen
  \bibfield  {author} {\bibinfo {author} {\bibfnamefont {S.}~\bibnamefont
  {Deffner}},\ }\bibfield  {title} {\bibinfo {title} {Geometric quantum speed
  limits: a case for wigner phase space},\ }\href
  {https://doi.org/10.1088/1367-2630/aa83dc} {\bibfield  {journal} {\bibinfo
  {journal} {New Journal of Physics}\ }\textbf {\bibinfo {volume} {19}},\
  \bibinfo {pages} {103018} (\bibinfo {year} {2017})}\BibitemShut {NoStop}%
\bibitem [{\citenamefont {O'Connor}\ \emph {et~al.}(2021)\citenamefont
  {O'Connor}, \citenamefont {Guarnieri},\ and\ \citenamefont
  {Campbell}}]{Connor2023}%
  \BibitemOpen
  \bibfield  {author} {\bibinfo {author} {\bibfnamefont {E.}~\bibnamefont
  {O'Connor}}, \bibinfo {author} {\bibfnamefont {G.}~\bibnamefont
  {Guarnieri}},\ and\ \bibinfo {author} {\bibfnamefont {S.}~\bibnamefont
  {Campbell}},\ }\bibfield  {title} {\bibinfo {title} {Action quantum speed
  limits},\ }\href {https://doi.org/10.1103/PhysRevA.103.022210} {\bibfield
  {journal} {\bibinfo  {journal} {Phys. Rev. A}\ }\textbf {\bibinfo {volume}
  {103}},\ \bibinfo {pages} {022210} (\bibinfo {year} {2021})}\BibitemShut
  {NoStop}%
\bibitem [{\citenamefont {Chitambar}\ and\ \citenamefont
  {G.}(2019)}]{Chitambar2019}%
  \BibitemOpen
  \bibfield  {author} {\bibinfo {author} {\bibfnamefont {E.}~\bibnamefont
  {Chitambar}}\ and\ \bibinfo {author} {\bibfnamefont {G.}~\bibnamefont {G.}},\
  }\bibfield  {title} {\bibinfo {title} {Quantum resource theories},\ }\href
  {https://doi.org/10.1103/RevModPhys.91.025001} {\bibfield  {journal}
  {\bibinfo  {journal} {Reviews of Modern Physics}\ }\textbf {\bibinfo {volume}
  {91}},\ \bibinfo {pages} {025001} (\bibinfo {year} {2019})}\BibitemShut
  {NoStop}%
\end{thebibliography}%
\end{document}